\documentstyle[aps,prb,psfig]{revtex}

\begin{document}

\renewcommand{\topfraction}{0.9}
\renewcommand{\bottomfraction}{0.9}
\renewcommand{\textfraction}{0.1}
\setcounter{topnumber}{3}
\setcounter{bottomnumber}{3}
\setcounter{totalnumber}{3}

\draft
\title{Conductivity in a symmetry broken phase:\\
 Spinless fermions with $1/d$ corrections}

\author{G\"otz S.~Uhrig\thanks{e-mail: uhrig@solrt.lps.u-psud.fr}}
\address{Laboratoire de Physique des Solides, B\^at.\ 510,
	Universit\'e Paris-Sud,
	F-91405 Orsay, France}

\date{\today}

\maketitle

\begin{abstract}
The dynamic conductivity $\sigma(\omega)$ of strongly correlated electrons
in a symmetry broken phase is investigated in the present work.
The model considered consists of spinless fermions with repulsive
interaction on a simple cubic lattice. The investigated symmetry broken
phase is the charge density wave (CDW) with wave vector 
${\bf Q}=(\pi,\pi,\pi)^\dagger$ which occurs at half-filling.
The calculations are based on the high dimensional approach, i.e.\
an expansion in the inverse dimension $1/d$ is used. The finite
dimensionality is accounted for by the inclusion of 
linear terms in $1/d$ and
the true finite dimensional DOS. Special care is paid to the
setup of a conserving approximation in the sense of Baym/Kadanoff
without inconsistencies. The resulting Bethe-Salpeter equation
is solved for the dynamic conductivity in the non symmetry broken and
in the symmetry broken phase (AB-CDW).
The dc-conductivity is reduced drastically in the CDW. Yet it does not
vanish in the limit $T\to 0$ due to a subtle cancellation of
 diverging mobility and vanishing DOS.
In the dynamic conductivity $\sigma(\omega)$ the energy gap
induced by the symmetry breaking is clearly discernible.
In addition, the vertex corrections of order $1/d$ lead to an excitonic 
resonance lying within the gap.
\end{abstract}
\pacs{71.27.+a, 71.30.+h, 71.45.Lr, 72.10.-d}

\section{Introduction}
The investigation of the transport properties of highly
correlated fermionic systems has attracted much
attention in recent years. A thorough understanding
of the conductivity in particular is essential
for the technical application of materials such
as metallic oxides in electronic devices.
The development of a new analytic
 approach, the limit of infinite dimension for fermionic
 systems \cite{metzn89a,vollh93},
allowed the numerical description of the metal-insulator
occuring in the half-filled Hubbard model in $d=\infty$ for higher
values of the interaction $U$ assuming a homogeneous phase
 \cite{georg96,prusc96}.
The latter assumption means that one deliberately ignores
the possible occurence of symmetry breaking for the sake
of simplicity. It is argued that on frustrated lattices
symmetry breaking is suppressed so that the metal-insulator
transition occurs at higher temperatures than those at which
 symmetry breaking sets in.

With this background in mind, it is the aim of this work
to extend and to complement the results known so far
 into two directions.
First, the finite dimensionality of realistic systems, i.e.\
mostly $d=3$, shall be included at least to lowest non-trivial
order in an expansion in $1/d$.
Much care is used in including these correction without
physical and/or analytic inconsistencies.
It is shown
that it is {\em not} sufficient to use a conserving, $\Phi$-derivable
approximation in the sense of Baym/Kadanoff.
 Furthermore, the true
 three-dimensional DOS will be used.
Second, the influence of symmetry breaking on the conductivity,
especially the question of possible metal-insulator transitions
induced by symmetry breaking shall be investigated.

To this end, the model of spinless fermions with repulsive
interaction for particles on adjacent sites is considered on
a generic bipartite lattice, namely the simple cubic lattice.
Its Hamiltonian at half-filling $n=1/2$ reads
\begin{equation}\label{hamil4}
\hat{H} = -\frac{t}{\sqrt{Z}}\sum_{<i,j>} {\hat c}^+_i{\hat
c}^{\phantom{+}}_j + \frac{U}{2Z}\sum_{<i,j>} {\hat n}_i {\hat n}_j
-\frac{U}{2} \sum_i {\hat n}_i\ .
\end{equation}
where $ {\hat c}^+_i \; ({\hat
c}^{\phantom{+}}_i)$ creates (annihilates) a fermion at site
$i$. The sum $\sum_{<i,j>}$ runs over all sites $i$ and $j$
which are nearest neighbors. The coordination number
$Z=2d=6$ appears for the proper scaling of
the kinetic energy \cite{metzn89a} and for the proper scaling of
the potential energy \cite{mulle89a}. The interaction constant
is $U$.

In this model the symmetry is broken yielding an AB-CDW at
half-filling \cite{halvo94} for infinitesimal values of the
interaction at $T=0$ and for sufficiently large interaction
at all finite temperatures. The AB-CDW consists of alternating
sites with a particle density above (below) average. The order
parameter $b$ is the absolute
 deviation of the particle density from its
average \cite{halvo94}.
As far as the occurence of a symmetry broken phase is concerned,
the model of spinless fermions at half-filling is similar
to the Hubbard model at half-filling which displays
 antiferromagnetic behavior. The main differences are that the
broken symmetry for spinless fermions is discrete whereas it
is continuous in the Hubbard model, and the fact that a local
interaction like the one in the Hubbard model does not
favor a spatial order by itself. The latter fact leads to a value
of $T_c \propto 1/U$ for large $U$ in the Hubbard model whereas
one has  $T_c \propto U$ in the spinless fermions model.

The article is organized as follows. Succeeding this introduction
it is discussed how a thermodynamically and analytically consistent
extension of the limit $Z\to \infty$ can be performed.
Next the basic equations for the extension to linear order
 $1/Z$ are derived and their
numerical evaluation is sketched.
 This third section contains also
results for the DOS and the corresponding proper self-energy.
 In sect.\ 4 the Bethe-Salpeter equation is set up
and solved for the conductivity $\sigma(\omega)$. The preservation
of the f-sum rule is discussed.
Numerical results for the dc- and the ac-conductivity
are presented in sect.\ 5.
The findings are
summarized and dicussed in the final section.

All energies (temperatures, respectively)
throughout this article will be given in units of the
root-mean-square of the ``free'', i.e.\ non-interacting,
density-of-states of the lattice model concerned.
All conductivities will be given in units
of $e^2/(\hbar a^{d-2})$ where $a$ is the lattice constant.
The constants $a$, $\hbar$, and $k_{\scriptstyle\rm B}$
(Boltzmann's constant) are set to unity.

\section{Proper self-consistent extension of $Z=\infty$}

In the case $Z=\infty$, the evaluation of diagrams and the treatment
of quantities like the DOS is conceptually simple. It is always
the leading contribution in $1/Z$ and only this which must be kept.
There is no dependence on the sequence in which certain quantities
and the equations relating them are considered. All sum rules
which hold in any dimension also hold at $Z=\infty$, continuity
provided for the limit $Z\to\infty$. This simplicity is lost
as soon as corrections in $1/Z$ are to be included. For
concreteness, let us consider the linear corrections $1/Z$; the
problems are illustrated for the free DOS, the Dyson equation
and the free energy $F$ as function of the order parameter $b$.

\begin{figure}[hbt]
\setlength{\unitlength}{1cm}
\begin{picture}(16,7)(0,0.7)
\put(8.3,0){\psfig{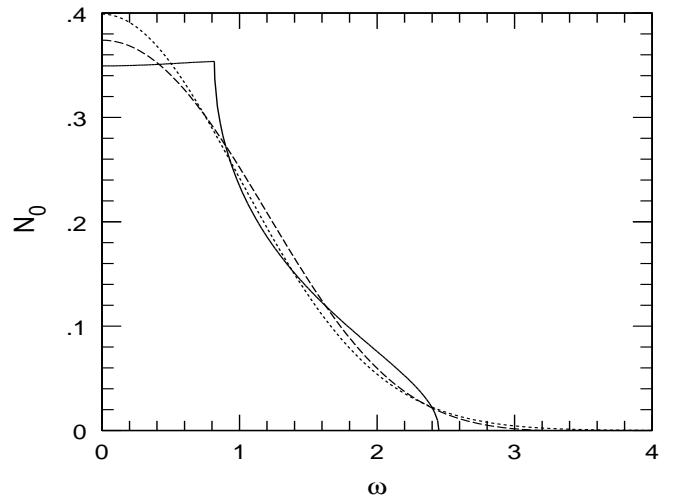}}
\end{picture}
\caption{Non-interacting DOS in $d=\infty$ 
	(short-dashed curve), in $d=3$
	(solid curve) and the DOS expanded in $1/d$ evaluated in $d=3$
	(long-dashed curve). These densities of states are symmetric
	about the y-axis.}
	\label{fi:1}
\end{figure}

The DOS is a non-negative function  of which the zeroth moment is
unity. This holds in any dimension, hence in $Z=\infty$. On including
 the linear corrections \cite{mulle89a} one realizes that the
approximate expression becomes negative at large values of $\omega$.
This is a disadvantage of the otherwise systematic expansion.
Another inconvenience catches the eye in fig.\ \ref{fi:1}.
The
expanded DOS does not improve considerably the agreement with the true
finite dimensional DOS (here $d=3$). A finite expansion in $1/Z$ cannot
produce the van-Hove-singularities.

To circumvent the problem of the DOS expansion, we decide to use
the exact finite dimensional DOS, i.e. the $d=3$ DOS.
This procedure provides often even in $d=1$ a remarkable
agreement \cite{halvo94,strac91}. In $d=3$, this approximation yields
qualitatively agreement for the local DOS as compared to finite
dimensional perturbation results \cite{schwe91a}.
Presently, the approach of using a finite dimensional DOS in an otherwise
infinite dimensional calculation as approximation for the finite
dimensional problem is employed as so-called ``dynamical mean-field
theory'' \cite{prusc96} or
``local impurity self-consistent approximation'' \cite{georg96}.

Next the problem of a systematic $1/Z$-expansion is discussed for
the Dyson equation. It is stated in a simple case when the
self-energy is strictly local in real space, i.e. constant in
momentum space
\begin{equation}\label{dysgl}
g(\omega)= g_0(\omega-\Sigma(\omega)) \ .
\end{equation}
This case is realized, for instance, in the
Hubbard model in $d=\infty$ \cite{mulle89a,janis92a}.
 No lattice site or spin index appears since the phase is assumed to be
homogeneous and non-magnetic. The quantity $g(\omega)$ stands for the
full local Green function $G_{i,i}(\omega)$ and $g_0(\omega)$
stands for the free Green function $G_{0;\, i,i}(\omega)$. The
expansion of the Green function corresponds to the expansion of the
thermodynamic potential since they depend linearly on each other
\cite{ricka80}. An expansion of the self-energy, however,
yields a {\em different} expression for $g(\omega)$ since $g_0(\omega)$
is not a linear function.
The expansion of the self-energy seems more
promising since it preserves the Dyson equation by construction.
Moreover, it is able to describe the shift of singularities, e.g. the
shifts of the band edges. (Note that we discuss here finite expansions of
the quantities considered).

In spite of the choice to expand the self-energy some ambiguity
persists. 
\begin{figure}[hbt]
\setlength{\unitlength}{1cm}
\begin{picture}(8.2,7)(0,0.7)
\put(8.3,0){\psfig{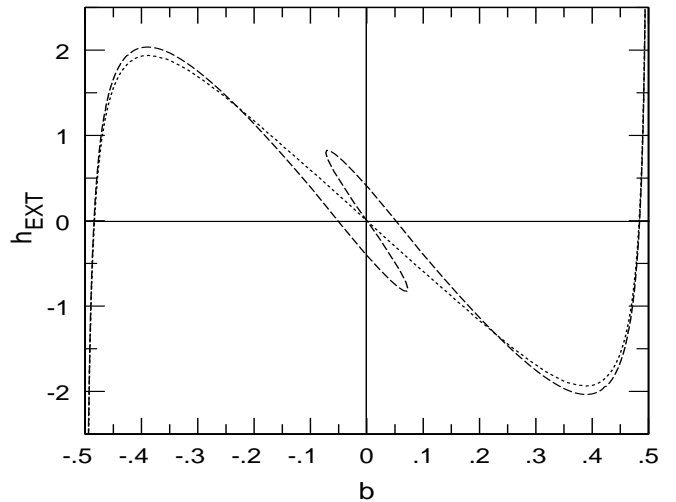}}
\end{picture}
\caption{Externally applied field $h_{\scriptstyle \rm EXT}$ 
	as function of the order
	parameter $b$ for $U=9$ and $T=0$ in $d=3$. The short-dashed
	curve depicts the $1/d$ self-consistent result, the long-dashed
	curve the result of a systematic expansion of the self-energy.
	The zeros of the curves correspond to thermodynamic equilibrium.
	But only zeros with positive slope are locally stable 
	($b\approx 0.48$).}
	\label{fi:2}
\end{figure}
In fig.\ \ref{fi:2}, this problem is illustrated.
It arises in the description of spontaneous symmetry breaking.
Two results for the
dependence of the conjugated field on the order parameter
are opposed. The data refers to the AB-CDW occuring in the spinless
fermion problem at half-filling. The dotted curve results from
a fully self-consistent calculation whereas the dashed curve results
from a systematic expansion of the self-energy. Note that
the self-consistent approach generates higher order contributions.

The argument results now from the strange behavior of the dashed curve
in the vicinity of the origin. The free energy belonging to
the dotted curve can be found by integration; it has an unstable maximum
($\partial h_{\scriptstyle\rm EXT}/\partial b <0$) at $b=0$ and
two stable minima ($\partial h_{\scriptstyle\rm EXT}/\partial b >0$)
at $b\approx \pm 0.48$. But there is no free energy belonging to the
dashed curve since it would have three maxima in sequence around
$b=0$ which is mathematically impossible (theorem of Rolle).
This is a very strong argument in favor
 of a self-consistent calculation.

For completeness, it shall be mentioned that one may argue that
in the vicinity of the physical solutions, i.e. the minima, the
difference of both approaches is negligible. There are also cases
known where the systematic, non self-consistent approach yields
better results \cite{schwe90b}. But there is still another advantage of
the self-consistent treatment
which will be crucial for what follows.
In the sense of Baym/Kadanoff \cite{baym61,baym62}
it covers also the calculations of two-particle properties and
ensures the preservation of sum rules.
So, Schweitzer and Czycholl resorted in their calulation of
resistance and thermopower for the periodic Anderson model to
the self-consistent treatment \cite{schwe91b}
 although their results for the local
DOS did not necessarily favor this approach \cite{schwe90b}.

\begin{figure}[hbt]
\setlength{\unitlength}{1cm}
\begin{picture}(8.2,4.2)(0,0)
\put(3,0.6){\psfig{file=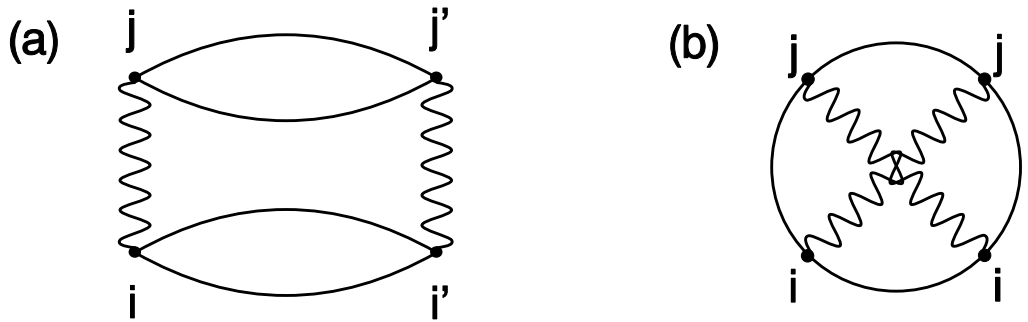,height=4.2cm,width=13.5cm}}
\end{picture}
\caption{Two examples of diagrams
	 in higher order (here: quadratic) in $1/Z$.
	The sites $i$ and $j$ are adjacent as are the sites $i'$ and $j'$.
	Additionally, $i\neq i'$ and $j\neq j'$ holds.}
	\label{fi:4}
\end{figure}

As result of the above discussion the starting point for the
inclusion of $1/Z$  correction is the generating functional
$\Phi$ according to Baym/Kadanoff \cite{baym61,baym62}. This
is the quantity which is expanded in a $1/Z$ series. Then the truncation
of this series yields an approximation to the corresponding order.
The power counting for the diagrams of $\Phi$ has been explained
previously \cite{halvo94,uhrig95d}. 
\begin{figure}[hbt]
\setlength{\unitlength}{1cm}
\begin{picture}(8.2,4.3)(0,0)
\put(2,0.6){\psfig{file=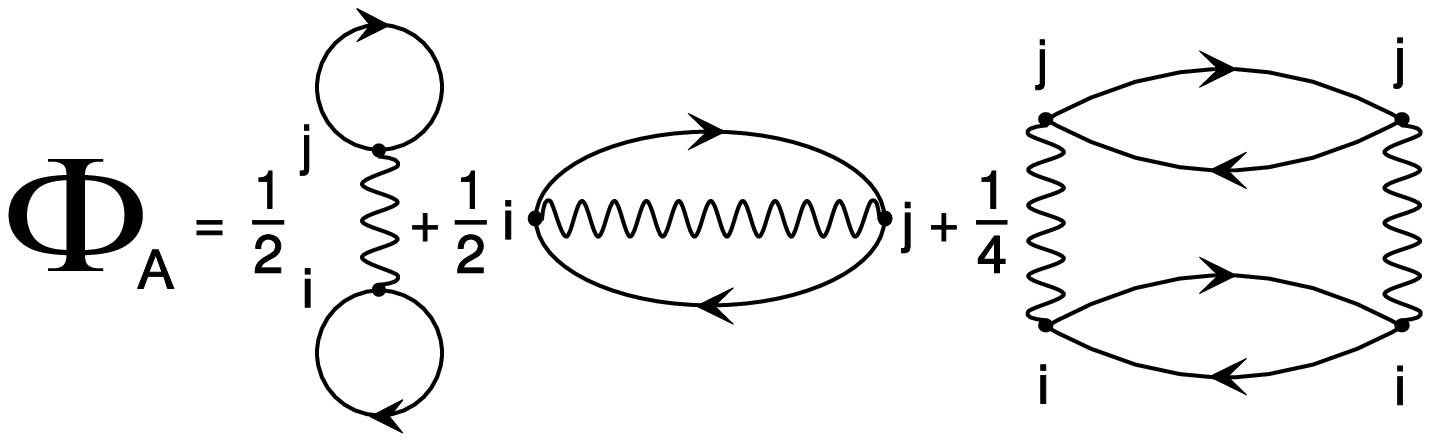,height=4.3cm,width=14.5cm}}
\end{picture}
\caption{Diagrams contained in 
	$\Phi_{\scriptstyle \rm A}[G]$. The first generates
	the Hartree term, the second the Fock term, and the third the
	local correlation term. The solid lines represent dressed
	propagators, the wavy lines the interactions. The sum runs over
	the lattice sites $i,j$.}
	\label{fi:3}
\end{figure}
Here it shall just be stated
that the first diagram in fig.\ \ref{fi:3} is of order ${\cal O}(1)$
and the two other diagrams in fig.\ \ref{fi:3} produce the linear
corrections ${\cal O}(1/Z)$ whereas the diagrams in
fig.\ \ref{fi:4} 
are examples for ${\cal O}(1/Z^2)$ contributions.
Thus fig.\ \ref{fi:3} visualizes the approximate 
$\Phi_{\scriptstyle\rm A}$ potential
which will be used in this work.

\begin{figure}[hbt]
\setlength{\unitlength}{1cm}
\begin{picture}(8.2,3.6)(0,0)
\put(4,0.6){\psfig{file=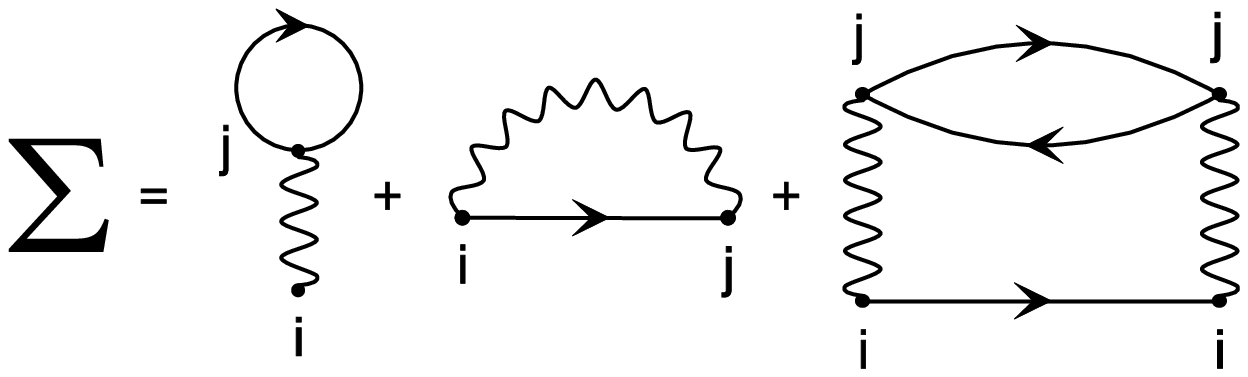,height=3.6cm,width=12.5cm}}
\end{picture}
\caption{The self-energy diagrams 
	derived from fig.\ \protect\ref{fi:3} by taking
	out one propagator line. The diagrams shown contribute in order
	$1/Z$.}
	\label{fi:5}
\end{figure}

By functional derivation the self-energy  shown in fig.\ \ref{fi:5}
is obtained.
Note that the Fock diagram is seemingly of another
order, namely ${\cal O}(1/Z^{3/2})$, than the third diagram,
 ${\cal O}(1/Z)$, which is called the local correlation
 diagram henceforth. What matters, however,
is the order relative to the free Green
 function which is ${\cal O}(1/Z^{1/2})$ for adjacent sites.
It is another advantage of the Baym/Kadanoff formalism that
one does not need to bother about these questions once the
approximate $\Phi$-potential is chosen.

Now a point shall be highlightened which
has not been mentioned before to our knowledge. In spite of the many
arguments in favor of the Baym/Kadanoff formalism its naive
application does not guarantee the absence of unphysical results.
A counter example serves as illustration. Consider an approximate
$\Phi$ consisting only of the diagram in fig.\ \ref{fi:4}(a),
 summed over all sites $i,j,i',j'$, such that $i$ and $j$
($i'$ and $j'$, respectively) are adjacent to one another and
fulfill $j\ne j'$ and $i\ne i'$. The resulting nearest-neighbor
self-energy $\Sigma_{i,j}=t(\omega)=t'(\omega)+i t''(\omega)$ has a finite
imaginary part $t''(\omega)$. Using the Dyson equation, one obtains
in the homogeneous phase
\begin{eqnarray}\nonumber
G_{{\bf k},{\bf k}}(\omega+0i) &=&
\frac{1}{\omega+0i-(1+t(\omega+0i))\varepsilon({\bf k})}
\\ \nonumber
&=& \frac{\omega-(1+t')\varepsilon({\bf k})+i t''\varepsilon({\bf k})}
{\left(\omega+0i-(1+t')\varepsilon({\bf k})\right)^2+
 \left(t''\varepsilon({\bf k})\right)^2}\ .
\end{eqnarray}
By choosing an appropriate wave vector ${\bf k}$ at fixed $\omega$
one can have the sign of $\varepsilon({\bf k})$ such that the imaginary
part of $G_{{\bf k},{\bf k}}(\omega+0i)$ is positive \cite{note1}. This is
a contradiction to the exact result \cite{ricka80}. Note that the
details of $t(\omega)$ are not essential as long as the imaginary
part is finite.

The counter example above is not only of academic interest.
Schweitzer and Czycholl observed as well that the inclusion
of a nearest-neighbor self-energy leads to wrong signs of the
imaginary parts. They considered the $1/d$ expansion of a $U^2$
perturbation theory around Hartree-Fock for the Hubbard model
and the periodic Anderson model \cite{schwe91a,schwe90a}.
 They reached consistency by
including higher $1/d$ corrections (for $d=1$ up to 50 terms)
\cite{schwe91a}.
Problems with the analyticity (uniqueness) of the solution
occurred also in the first investigations
of $1/d$ corrections in the Hubbard model \cite{georg96}
(Falicov-Kimball model \cite{schil95}).

To the author's knowledge there is no necessary or sufficient
theory so far, which predicts under which circumstances
such problems have to be expected or can be excluded.
A sufficient argument excluding wrong signs of the imaginary
part of the approximate self-energy is given by the theorem:

If the approximation considered can be interpreted as an
expansion of the self-energy in a parameter $\lambda>=0$
and if $m$ is the leading order, in which the imaginary
part of the self-energy does {\em not} vanish, then
the self-energy approximated in the $m$-th order has the
right sign.

The proof relies on the continuity of limits if the expansion exists.
 According to the precondition holds
\begin{equation}\label{gegen2}
0 \geq {\mathop{\rm Im}\nolimits} \Sigma_{\lambda}(\omega,{\bf k})
\, = \, \lambda^m {\mathop{\rm Im}\nolimits} \Sigma^{(m)}(\omega,{\bf k})
+{\cal O}(\lambda^{(m+1)}) \ ,
\end{equation}
which is equivalent to
\begin{equation}\label{gegen3}
0 \geq \lim_{\lambda\to 0+} \lambda^{-m}
{\mathop{\rm Im}\nolimits}\Sigma_{\lambda}(\omega,{\bf k})
\, = \, {\mathop{\rm Im}\nolimits} \Sigma^{(m)}(\omega,{\bf k})\ .
\end{equation}
The index ${\bf k}$ is the wave vector in a homogeneous,
translationally invariant phase. The derivation for
 general phases, for instance
the AB-CDW, is given in appendix A.

The derivation in (\ref{gegen2}) and in (\ref{gegen3}) holds strictly
only for the non self-consistent treatment. In the generic situation,
however, the leading order of the self-energy with non-vanishing
imaginary part results from a certain diagram class and the analytic
properties do not depend on the specific form of the Green function
entering. If this is the case, the statement of the theorem extends
also to the self-consistent treatment where the quantitative form of the
Green functions are not known a priori.

The theorem helps one to understand the observations made
by Schweitzer and Czycholl. In the $1/d$ expansion of the
 $d$-dimensional Hubbard model and of the periodic Anderson model
one has $\lambda=1/d$ and $m=0$ since the self-energy is imaginary
already in the first order. For the perturbation theory in $U$
one has $\lambda=U$ and $m=2$ since the self-energy stays real
in Hartree-Fock. Applying the rationale  of the theorem
twice one understands that the self-energy in $U^2$ of the infinite
dimensional model has the right analytic behavior. If further
$1/d$ corrections are included this does not need to be true.
The result of Schweitzer and Czycholl, that the linear $1/d$ correction
leads to wrong signs, proves that the theorem is sharp: If the
precondition fails, the implication fails, too. The second obervation,
that the inclusion of {\em very many} $1/d$ correction terms remedies
the failure, can also be understood easily. In this case the
calculations approximates the $U^2$ perturbation theory of the
{\em finite} dimensional models very well. According to the theorem,
this perturbation theory displays the right sign, too.

The above observations indicate that also the analyticity problems
encountered for $1/d$ corrections in the Hubbard model \cite{georg96}
are not due to the approximations used to solve the effective
impurity problems. Rather each time that the theorem does not 
apply one has to expect that analyticity problems arise for certain
parameters.
Considering eq.\ (6a) in ref.\ 18 or equivalently eq.\ (370) in ref.\ 3
one realizes that the spectral density of the local self energy might
change sign. This cannot obviously be excluded from the way how the
impurity self energies are computed.

Turning to the $1/d$ expansion of
the present model of spinless fermions ($\lambda=1/d$), one notes that
the theorem applies with $m=1$. Therefore, the equations
including linear $1/d$ corrections display the right analyticity.
These equation will be set up in the following.

\subsection{Resulting equations and one-particle results}
This section is kept very concise since it contains
material which is partly published elsewhere \cite{halvo94}.
For two reasons, however, it cannot be omitted. Firstly,
a different notation using different intermediate quantities
shall be introduced. Secondly, the one-particle results
are necessary requisites to understand the conductivity
results in the subsequent section.

The treatment of a self-energy of the type depicted in
fig.\ \ref{fi:5} is commonly known (see e.\ g.\ refs.\
11, 19, 20). Dealing with the symmetry
broken phase, however, requires some extension. In a previous work
\cite{halvo94} local Green function and the self-energy are distinguished
according to the sublattice to which they belong. In the present work,
 sum and difference of the quantities on the two sublattices
will be used. The local quantities on site $i$ belonging to
sublattice $\tau \in \{A,B\}$ are
\begin{mathletters}
\label{sigdef1}
\begin{eqnarray}\label{sigdef1a}
g_\tau &:=& G_{i,i}(\omega)
\\ \label{sigdef1b}
\Sigma_\tau(\omega)&:=&\Sigma^{\scriptstyle\rm H}_{i,i}(\omega)
+\Sigma^{\scriptstyle\rm C}_{i,i}(\omega)  \ ,
\end{eqnarray}
\end{mathletters}
where $G_{i,i}(\omega)$ is the full local Green function
and $\Sigma$ is the local self-energy.
The Fock part will be treated subsequently. The index $^{\scriptstyle\rm H}$
stands for the Hartree term (first diagram in fig.\ \ref{fi:5});
 the index $^{\scriptstyle\rm C}$ stands for the local correlation
(third diagram in fig.\ \ref{fi:5}). Let us define
\begin{mathletters}
\label{sumdif}
\begin{eqnarray}\label{sumdifa}
g_{\scriptstyle\rm S}(\omega)&:=&
(g_{\scriptstyle\rm A}(\omega)+g_{\scriptstyle\rm B}(\omega))/2
\\ \label{sumdifb}
g_{\scriptstyle\rm D}(\omega)&:=&
(g_{\scriptstyle\rm A}(\omega)-g_{\scriptstyle\rm B}(\omega))/2
\\ \label{sumdifc}
\Sigma(\omega)&:=&(\Sigma_{\scriptstyle\rm A}(\omega)+
\Sigma_{\scriptstyle\rm B}(\omega))/2
\\ \label{sumdifd}
\Delta(\omega)&:=&(\Sigma_{\scriptstyle\rm A}(\omega)-
\Sigma_{\scriptstyle\rm B}(\omega))/2 \ .
\end{eqnarray}
\end{mathletters}
The spectral functions of the Green function
are called $N_{\scriptstyle\rm S}$ and $N_{\scriptstyle\rm D}$, respectively;
the spectral functions of the self-energy $\Sigma$ and $\Delta$
are called $N_\Sigma$ and $N_\Delta$, respectively.
The non local Fock term is $\Sigma^{\scriptstyle\rm F}:=\Sigma_{i,j}$, where
 $i$ and $j$ are adjacent sites. It turns out, that 
$\Sigma^{\scriptstyle\rm F}$
is negative (for repulsive interaction),
 real, and that it does not depend on whether the fermion hops
from $A$ to $B$ or vice versa. Hence, it renormalizes the hopping
\begin{equation}\label{tren}
t \to \gamma t \quad \mbox{with} \quad \gamma:=1-\sqrt{Z} 
\Sigma^{\scriptstyle\rm F}/t\ .
\end{equation}
Note that for attractive interaction $\gamma$ could become 0 which
would lead to a breakdown of the theory. Such a singularity is
absent in the repulsive case.

In the  AB-CDW, the modes at ${\bf k}$ couple to those at
${\bf k}+{\bf Q}$. Hence one has
\begin{equation}\label{block2}
\left( \begin{array}{lr}
G_{{\bf k},{\bf k}}&  G_{{\bf k},{\bf k}+{\bf Q}}\\
G_{{\bf k+{\bf Q}},{\bf k}} & G_{{\bf k+{\bf Q}},{\bf k}+{\bf Q}}
 \end{array} \right)
= \left( \begin{array}{lr}
	\omega-\Sigma(\omega)-\gamma\varepsilon & -\Delta(\omega) \\
	-\Delta(\omega)   & \omega-\Sigma(\omega)+\gamma
	 \varepsilon \end{array} \right)^{-1} \ .
\end{equation}
From this equation one obtains
\begin{mathletters}
\label{green3}
\begin{eqnarray}\nonumber
g_{\scriptstyle\rm S}(\omega)&=&
\frac{w}{\gamma\sqrt{w^2-\Delta^2(\omega)}}
g_0(\sqrt{w^2-\Delta^2(\omega)}/\gamma)
\\ \label{green3a}
&=&\int\limits_{-\infty}^\infty \frac{w}{w^2-(\gamma\varepsilon)^2-\Delta^2}
 N_0(\varepsilon)d\varepsilon
\end{eqnarray}
\begin{eqnarray}\nonumber
g_{\scriptstyle\rm D}(\omega)&=&
\frac{\Delta(\omega)}{\gamma\sqrt{w^2-\Delta^2(\omega)}}
g_0(\sqrt{w^2-\Delta^2(\omega)}/\gamma)
\\ \label{green3b}
&=&\int\limits_{-\infty}^\infty 
\frac{\Delta}{w^2-(\gamma\varepsilon)^2-\Delta^2}
 N_0(\varepsilon)d\varepsilon
\ , \end{eqnarray}
\end{mathletters}
where $w$  is short hand for $\omega-\Sigma(\omega)$.

The averaged Hartree term $U(n_{\scriptstyle\rm A}+n_{\scriptstyle\rm B})/2$
renormalizes the chemical potential \cite{uhrig95d}. The Hartree
contribution to $\Delta$ is $Ub$ where 
$b:=(n_{\scriptstyle\rm B}-n_{\scriptstyle\rm A})/2$
 is the order parameter, i.e.\ the particle density difference.
It is given by $b=-\int\limits_{-\infty}^\infty 
N_{\scriptstyle\rm D}(\omega) f_{\scriptstyle\rm F}(\omega) d\omega $,
where $f_{\scriptstyle\rm F}(\omega)$ is the Fermi function.

The Fock term can be calculated from the nearest-neighbor
Green function $G_{j+a,j}$
\begin{equation}\label{kinen2}
\Sigma^{\scriptstyle\rm F} = \frac{U}{\pi Z}
\int\limits_{-\infty}^\infty 
{\mathop{\rm Im}\nolimits}\left( G_{j+a,j}(\omega+0i)
 \right) f_{\scriptstyle\rm F}(\omega)d\omega \ ,
\end{equation}
which is given by
\begin{eqnarray}\nonumber
G_{j+a,j}(\omega)&=& -\frac{1}{\sqrt{Z}} \int\limits_{\scriptstyle\rm BZ}
\varepsilon({\bf k})G_{{\bf k},{\bf k}} \frac{dk^d}{(2\pi)^d}
\\ \label{kinen3}
&=& -\frac{1}{\gamma\sqrt{Z}}
  \left[(\omega-\Sigma) g_{\scriptstyle\rm S}(\omega)-
\Delta g_{\scriptstyle\rm
D}(\omega)\right]
 d\varepsilon \ .
\end{eqnarray}
The Fock term is related to the kinetic energy
$\Sigma^{\scriptstyle\rm F} =(U/Z^{3/2})\langle \hat T \rangle$.
Thus, (\ref{kinen2}) can be evaluated using (\ref{kinen3}) and (\ref{green3}).

The local correlation term is given in terms of the Matsubara frequencies
$\omega_\lambda$ (fermionic) and $\omega_l$ (bosonic) by
\begin{equation}\label{korr0}
\Sigma_\tau^{\scriptstyle\rm C}(i\omega_\nu) = -\frac{U^2T^2}{Z}
\sum\limits_{l,\lambda} g_{\overline{\tau}}(i\omega_\lambda+i\omega_l)
g_{\overline{\tau}}(i\omega_\lambda)g_\tau(i\omega_\l+i\omega_\nu)
\ .
\end{equation}
Here, the index $\overline{\tau}$ stands for the {\em other}
sublattice, i.e.\ for $A$ if $\tau=B$ and vice versa.
By performing the Matsubara sum one obtains the convolution
\begin{eqnarray}\nonumber
N_{\Sigma_\tau}(\omega)
& = &\frac{U^2}{Z}\int\limits_{-\infty}^\infty \int\limits_{-\infty}^\infty 
N_{\overline{\tau}}(\omega'') N_{\overline{\tau}}(\omega''-\omega')
  N_\tau(\omega-\omega') \cdot \hspace{0.5cm}
\\ && \hspace{0.5cm}  \label{korr4}
\left[ f_{\scriptstyle\rm F}(\omega'-\omega)
f_{\scriptstyle\rm F}(-\omega'')f_{\scriptstyle\rm F}(\omega''-\omega')+
f_{\scriptstyle\rm F}(\omega-\omega')
f_{\scriptstyle\rm F}(\omega'')
f_{\scriptstyle\rm F}(\omega'-\omega'') \right]
d\omega' d\omega''
\end{eqnarray}
for the spectral function $N_{\Sigma_\tau}(\omega)$
belonging to $\Sigma_\tau^{\scriptstyle\rm C}(\omega)$.
The convolution can be expressed most conveniently in the
Fourier transforms
\begin{mathletters}
\label{four1}
\begin{eqnarray}\label{four1a}
\widetilde N^\pm(t)&:=&\int\limits_{-\infty}^\infty
 \exp{(-i\omega t)} N(\pm\omega) f_{\scriptstyle\rm F}(-\omega)
\\ \label{four1b}
\widetilde N(t)&:=&\int\limits_{-\infty}^\infty \exp{(-i\omega t)} N(\omega)
\ .
\end{eqnarray}
\end{mathletters}
Eq.\ (\ref{korr4}) becomes as simple as $\widetilde N_{\Sigma_\tau}(t) =
\frac{U^2}{Z} \left[\left.\widetilde N_{\overline{\tau}}^+
 \widetilde N_\tau^+\widetilde N_{\overline{\tau}}^-\right|_t
+ \left.\widetilde N_{\overline{\tau}}^-\widetilde N_\tau^- 
\widetilde N_{\overline{\tau}}^+
\right|_{-t}\right]$. In sums and differences one obtains
\begin{mathletters}\label{four4}
\begin{eqnarray}
\label{four4a}
\widetilde N_{\Sigma}(t) &=&
\frac{U^2}{Z} \left[\left. \{(\widetilde N_{\scriptstyle\rm S}^+)^2-
(\widetilde N_{\scriptstyle\rm D}^+)^2\} 
\widetilde N_{\scriptstyle\rm S}^-\right|_t
+ \left. \{(\widetilde N_{\scriptstyle\rm S}^+)^2-
(\widetilde N_{\scriptstyle\rm D}^+)^2\} 
\widetilde N_{\scriptstyle\rm S}^+\right|_{-t}\right]
\\ \label{four4b}
\widetilde N_{\Delta}(t) &=&
-\frac{U^2}{Z} \left[\left. \{(\widetilde N_{\scriptstyle\rm S}^+)^2-
(\widetilde N_{\scriptstyle\rm D}^+)^2\} 
\widetilde N_{\scriptstyle\rm D}^-\right|_t
+\left. \{(\widetilde N_{\scriptstyle\rm S}^+)^2-
(\widetilde N_{\scriptstyle\rm D}^+)^2\} 
\widetilde N_{\scriptstyle\rm D}^+\right|_{-t}\right]
\ .\end{eqnarray}
\end{mathletters}
The complete self-energy $\Sigma$ and $\Delta$
are given by the following inverse Fourier transforms
\begin{mathletters}\label{four5}
\begin{eqnarray} \label{four5a}
\Sigma(\omega+0i)&=& -i\int\limits_0^\infty \exp{(i\omega t-0t)}
\widetilde N_{\Sigma}(t) dt
\\ \label{four5b}
\Delta(\omega+0i)&=& Ub -i\int\limits_0^\infty \exp{(i\omega t-0t)}
\widetilde N_{\Delta}(t) dt\ .
\end{eqnarray}
\end{mathletters}
In (\ref{four5b}) the Hartree part has been added.

So far, no assumptions concerning the DOS entered. The
formulae hold for all fillings. At the particular value
of half-filling the additional symmetries
$N_{\scriptstyle\rm S}(\omega)=N_{\scriptstyle\rm S}(-\omega)$,
$N_{\scriptstyle\rm D}(\omega)=-N_{\scriptstyle\rm D}(-\omega)$,
$N_{\Sigma}(\omega)=N_{\Sigma}(-\omega)$  and
$N_\Delta(\omega)=-N_{\Delta}(-\omega)$ can be
exploited.
The fact that the spectral densities
are real tells us that $\widetilde N(-t)$ is the
complex conjugate (c.c.) of $\widetilde N(t)$.
Thus (\ref{four4}) simplifies at half-filling to
\begin{mathletters}\label{four7}
\begin{eqnarray}\label{four7a}
\widetilde N_{\Sigma}(t) &=&
\frac{U^2}{Z} \left[\left. \{(\widetilde N_{\scriptstyle\rm S}^+)^2-
(\widetilde N_{\scriptstyle\rm D}^+)^2\} 
\widetilde N_{\scriptstyle\rm S}^+\right|_t
+{}\mbox{c.c.} \right]
\\ \label{four7b}
\widetilde N_{\Delta}(t) &=&
\frac{U^2}{Z} \left[\left. \{(\widetilde N_{\scriptstyle\rm S}^+)^2-
(\widetilde N_{\scriptstyle\rm D}^+)^2\} 
\widetilde N_{\scriptstyle\rm D}^+\right|_t
-{}\mbox{c.c.} \right]
\ .\end{eqnarray}
\end{mathletters}
This terminates the set up of the equations which have to be
solved self-consistently on the one-particle level.

For those who intend to implement these equations or similar ones
some remarks on the numerical realization are in order.
As usual, the self-consistent set of equations is solved by
iteration. At $T=0$ it is favorable to use a relaxed iteration.
This means that the self-energy $\Sigma$ and $\Delta$ from the
$n$-th and from the $n+1$ iteration are averaged and used
for the subsequent calculation instead of using only the $n+1$ iteration.
This procedure damps oscillatory deviations from the fixed point
more rapidly. It is even more advantageous to let the programme
decide whether relaxed or non relaxed iteration converges faster.

The Fourier transformation is the most time consuming step.
The best algorithm for this task is the so called Fast Fourier
Transformation (FFT). The extremely large number of points,
which can be used with the FFT, overcompensates the disadvantage
of an equidistant mesh which cannot be adapted to regions where
the  DOS changes
rapidly \cite{halvo94}.
In the AB-CDW $2^{19}$ points were used. The vectorization
on a IBM3090 still permitted to do one iteration step comprising
four FFT in 19 seconds. A very good precision could be achieved.
The sum rules
\begin{mathletters}\label{sum}
\begin{eqnarray}\label{suma}
\int\limits_0^\infty N_{\Sigma}(\omega)d\omega &=& \frac{U^2}{Z}\frac{1}{2}
\left(\frac{1}{4} -b^2\right)
\\ \label{sumb}
\int\limits_0^\infty N_{\Delta}(\omega)d\omega &=& \frac{U^2}{Z} b
\left(\frac{1}{4} -b^2\right)
\end{eqnarray}
\end{mathletters}
are preserved up to $10^{-6}$. Note that (\ref{sumb}) holds
only at $T=0$ whereas (\ref{suma}) holds for all temperatures.

In order to achieve the high precision also at $T=0$, it is
necessary to discretize the DOS carefully. At the gap edges
the DOS displays inverse square root divergences
$a/\sqrt{\omega-\omega_{\Delta}}$. The parameters $a$ and $\omega_{\Delta}$
are determined directly from the self-energy using (\ref{green3}).
The diverging part of the DOS is discretized by using the
 average value in the interval
$[\omega_i-\delta\omega/2,\omega_i+\delta\omega/2]$ instead of the
DOS value at $\omega_i$.

Once the Fourier transforms are essentially linear one as to avoid
a non-linear time loss
in the calculation of the complex free Green function
$g_0(z)$. Therefore, the integration from the Hilbert representation
must be avoided. This is done by using the approximate expression
\begin{eqnarray}
N_3(\varepsilon) &\approx&
\frac{1}{\pi} \left[
\left\{ \frac{13033}{29088}+\frac{8675}{174528}\varepsilon^2 \right\}
\sqrt{6-\varepsilon^2} - {} \right.
\nonumber\\
&& \left\{ \frac{4167}{6464} +\frac{459}{6464\sqrt{6}}\varepsilon
+ \frac{729}{12928}\varepsilon^2 \right\}
\sqrt{2/3-(\varepsilon-2\sqrt{2/3})^2} - {}
\nonumber\\
&&\left. \left\{ \frac{4167}{6464} -\frac{459}{6464\sqrt{6}}\varepsilon
+ \frac{729}{12928}\varepsilon^2 \right\}
\sqrt{2/3-(\varepsilon+2\sqrt{2/3})^2}
\right] \label{rho3}
\ ,\end{eqnarray}
for the three dimensional DOS $N_3(\varepsilon)$. The identities
$h(z;a):= (1/\pi)\int_{-\sqrt{a}}^{\sqrt{a}}
\sqrt{a-\varepsilon^2}/(z-\varepsilon) d\varepsilon =z\pm\sqrt{z^2-a}$
and $(1/\pi)\int_{-\sqrt{a}}^{\sqrt{a}}
\varepsilon \sqrt{a-\varepsilon^2}/(z-\varepsilon)
d\varepsilon = -(a/2)+ z h(z;a)$ permit to
compute $g_0(z)$ for any $z$ quickly.
The r.h.s.\ of (\ref{rho3}) is chosen such that the van-Hove-singularities
are at the right places and such that the first moments (including the 8th)
are reproduced exactly. The relative accuracy achieved is $4\cdot10^{-4}$
for $N_3(0)$ and $10^{-5}$ for the 10th and the 12th moment.

The calculation of the Hartree and of the Fock parts are linear
in the number of discretization points. Concluding the remarks
on the numerical realization we state that all parts of an
iteration step are essentially linear in the number of points used.
This allows a reliable and efficient computation.

\begin{figure}[hbt]
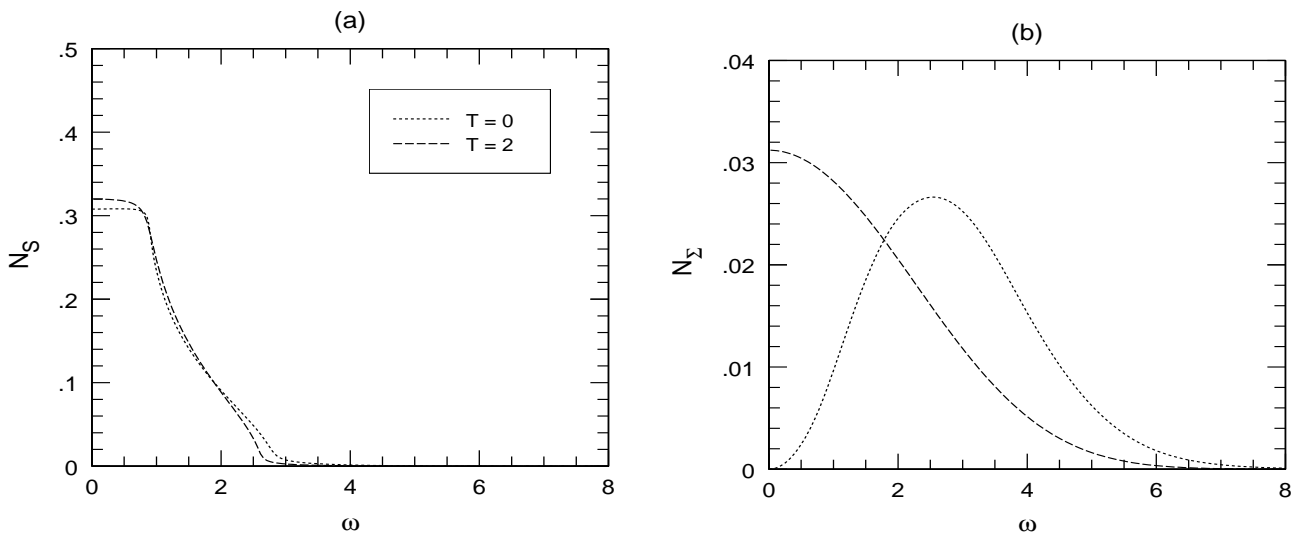

\setlength{\unitlength}{1cm}
\begin{picture}(8.2,7)(0,0.7)
\put(-0.7,0){\psfig{file=fig6a.ps,height=7.2cm,width=8cm,angle=270}}
\put(8.3,0){\psfig{file=fig6b.ps,height=7.2cm,width=8cm,angle=270}}
\end{picture}
\caption{Density of states and 
	spectral function of the self-energy in the
	homogeneous phase at $U=2$ and $T=0$ and $T=2$ in $d=3$. For
	definitions see eqs.\ (\protect\ref{sumdif}).}
	\label{fi:6}
\end{figure}
In fig.\ \ref{fi:6}, results for the DOS and the spectral density
of the self-energy in the homogeneous phase
are shown. The spontaneous symmetry breaking is deliberately
suppressed.
Only positive frequencies are displayed since the
functions are even. At $T=0$, one notes that the imaginary part
of the self-energy tends quadratically to zero for $\omega \to 0$.
From (\ref{korr4}) this follows for all free DOSes with finite
non-singular value at the Fermi edge. Thus the homogeneous low
temperature phase of interacting spinless fermions is a
Fermi liquid. But this phase is thermodynamically unstable (see
below). The DOS still bears signs of the van-Hove-singularities
which are smeared out only a little due to the interaction.
Note that the width is increased by the Fock term. In the free case
the half-width is $\sqrt{6} \approx 2.45$.
High temperatures smear out the minimum of $N_\Sigma$ at $\omega=0$
completely. The solution depicted is stable since at $T=2$
 no AB-CDW is possible.

\begin{figure}[hbt]
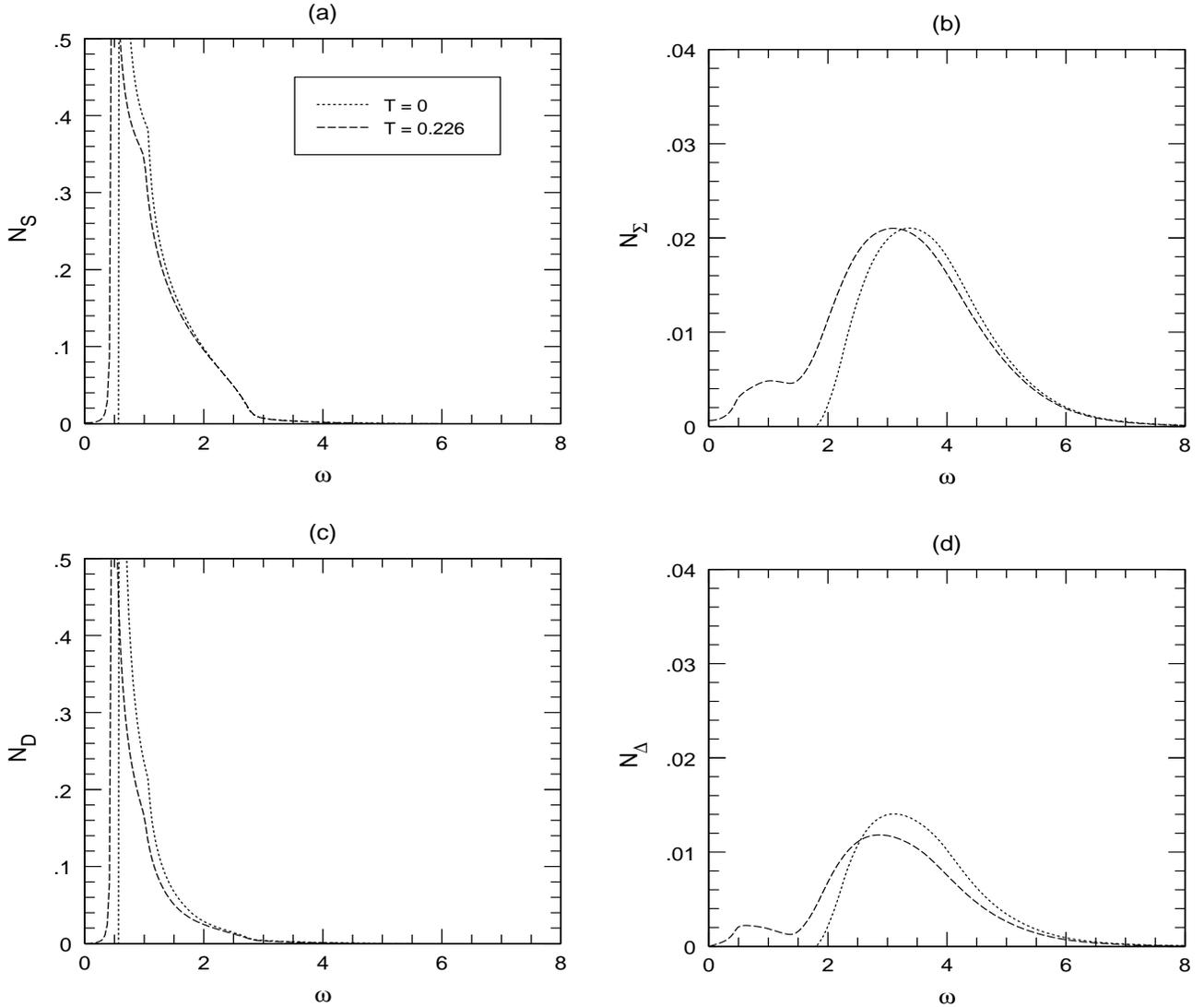

\setlength{\unitlength}{1cm}
\begin{picture}(8.2,15)(0,0.7)
\put(-0.7,7.5){\psfig{file=fig7a.ps,height=7.2cm,width=8cm,angle=270}}
\put(8.3,7.5){\psfig{file=fig7b.ps,height=7.2cm,width=8cm,angle=270}}
\put(-0.7,0){\psfig{file=fig7c.ps,height=7.2cm,width=8cm,angle=270}}
\put(8.3,0){\psfig{file=fig7d.ps,height=7.2cm,width=8cm,angle=270}}
\end{picture}
\caption{Density of states and
	 spectral function of the self-energy in
	the AB-charge density wave at $U=2$ and $T=0$ ($b=0.311005$),
	$T=0.225658$ ($b=0.250000$) in $d=3$. For
	definitions see eqs.\ (\protect\ref{sumdif}). The sum quantities in 
	(a) and (b) are even functions of frequency; the difference
	quantities in (c) and (d) are odd functions.}
	\label{fi:7}
\end{figure}
In fig.\ \ref{fi:7}, stable solutions with $b>0$ are shown. Note the
square root divergence in the DOSes (left column) in the vicinity
of the gap.
At $T=0$ the gap is at $2\omega_\Delta\approx0.6$
 whereas the spectral density of the self-energy
 becomes finite at about $1.8\approx 6\omega_\Delta$. This results
from the two convolutions involved \cite{halvo94}. They make the
gap in the density of the self-energy to be exactly three times
the gap in the DOS.
Put differently, the finite spectral
density of the self-energy corresponds to
the inelastic scattering of a particle or a hole involving
an additional particle-hole pair. Thus, the necessary minimum energy
is three times the elementary gap.
The physically important implication is the
existence of quasi-particles with energies
between $\omega_\Delta$ and $3\omega_\Delta$
with infinite life-time. Following the arguments of Luttinger
\cite{lutti61} by which he shows that the density of the
self-energy generically
goes like $\omega^2$ at the Fermi edge one comes to the conclusion
that this factor 3 is not an artifact of the approximation but
valid to all orders. Therefore, if the conditions are such that the
the homogeneous phase is a Fermi liquid, i.e.\ Luttinger's
argument holds, a gapped, spontaneously symmetry broken phase
has a factor 3 between the gap in the DOS and the gap in the
self-energy. This implies also the existence of undamped quasi-particles
which have interesting consequences on the transport properties
(see below). The exponent of the
power law with which the imaginary parts of the
self-energy rises at $\omega=3\omega_\Delta$ is $3/2$.

At finite temperatures the energy gap is smaller since the
order parameter has decreased. This effect is visible already
in the Hartree treatment. In addition, the energy gap is smeared out:
thermal fluctuations
represented by the local correlation term $\Sigma^{\scriptstyle\rm C}$
 induce a certain spectral weight within
the ``gap'' which does no longer exist in the rigorous sense.
The occurrence of two maxima in $N_\Sigma$ and in
$N_\Delta$ should be noted.

\begin{figure}[hbt]
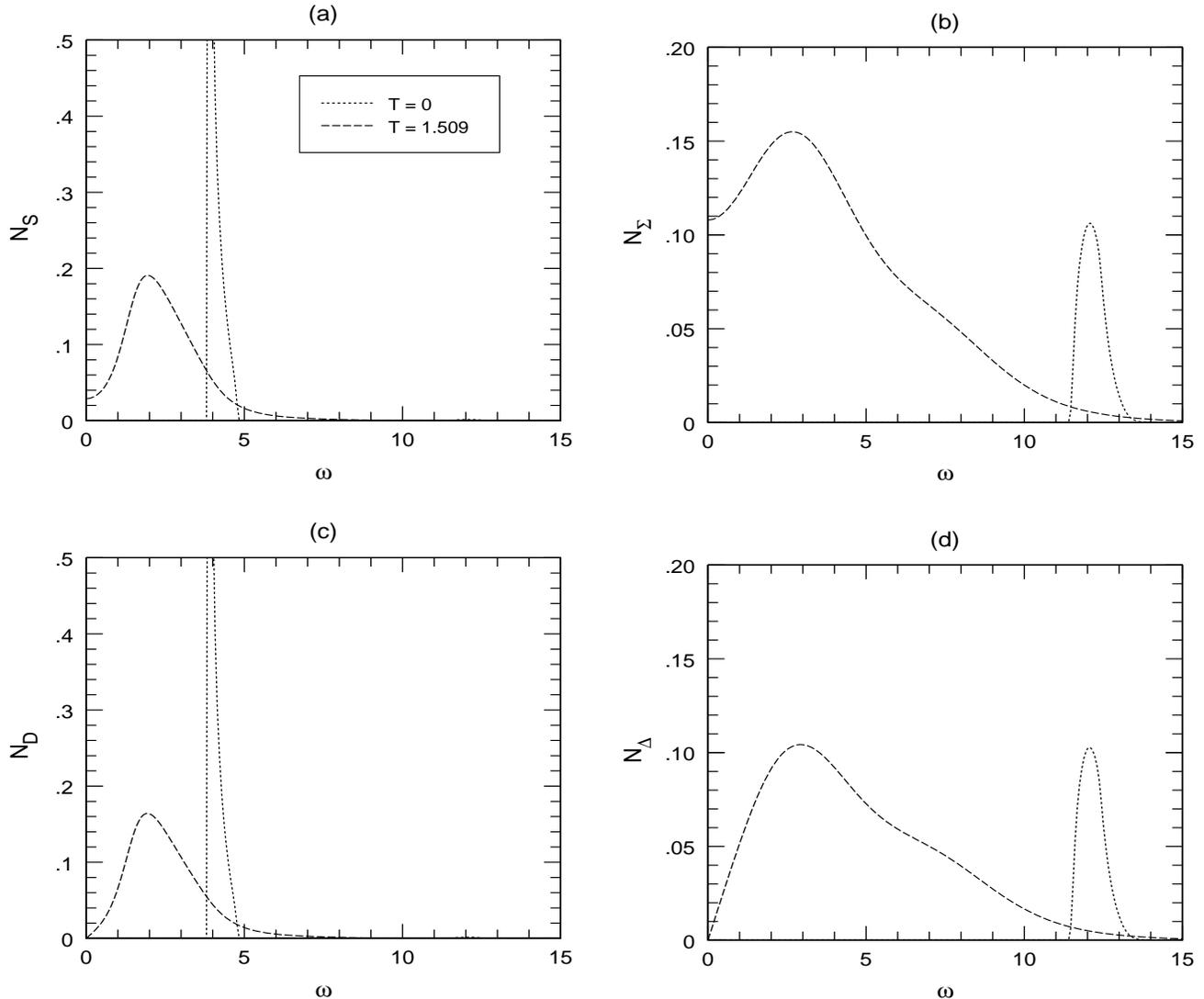

\setlength{\unitlength}{1cm}
\begin{picture}(8.2,15)(0,0.7)
\put(-0.7,7.5){\psfig{file=fig8a.ps,height=7.2cm,width=8cm,angle=270}}
\put(8.3,7.5){\psfig{file=fig8b.ps,height=7.2cm,width=8cm,angle=270}}
\put(-0.7,0){\psfig{file=fig8c.ps,height=7.2cm,width=8cm,angle=270}}
\put(8.3,0){\psfig{file=fig8d.ps,height=7.2cm,width=8cm,angle=270}}
\end{picture}
\caption{Density of states and 
	spectral function of the self-energy in
	the AB-charge density wave at $U=8$ and $T=0$ ($b=0.479312$),
	$T=1.509384$ ($b=0.260004$) in $d=3$. For
	definitions see eqs.\ (\protect\ref{sumdif}). 
	At $\omega \approx 12.0$ hardly visible satellite
	bands are present in $N_{\scriptstyle\rm S}$ and
	$N_{\scriptstyle\rm D}$ for $T=0$. They result from
	the imaginary parts of the self-energy around
	this frequency.}
	\label{fi:8}
\end{figure}
In fig.\ \ref{fi:8}, the generic results for large values
of the interaction are shown. At $T=0$ the factor $3$
between the gap in the DOSes and the gap of the spectral
densities of the self-energies is even more easily discernible.
At the finite temperature ($T\approx 1.5$), all the structures
are smeared out; the order parameter is considerably smaller
than at $T=0$: $b=0.260$ at finite $T$ to $b=0.479$ at $T=0$.
The comparison of the spectral weights of the self-energy at zero and at
finite temperature illustrates an important effect.
The correlation term is suppressed by the symmetry breaking.
The larger $b$ the smaller is the area under the curves
in fig.\ \ref{fi:8}(b) and (d). The effect can be understood
quantitatively with the help of the equations (\ref{sum})
which imply that the area under the curves vanishes for $b\to 1/2$.
This leads to the counter-intuitive
effect that the significance of the
correlation term decreases on increasing interaction at $T=0$
albeit it is quadratic in the interaction

\begin{figure}[hbt]
\setlength{\unitlength}{1cm}
\begin{picture}(8.2,7)(0,0.7)
\put(8.3,0){\psfig{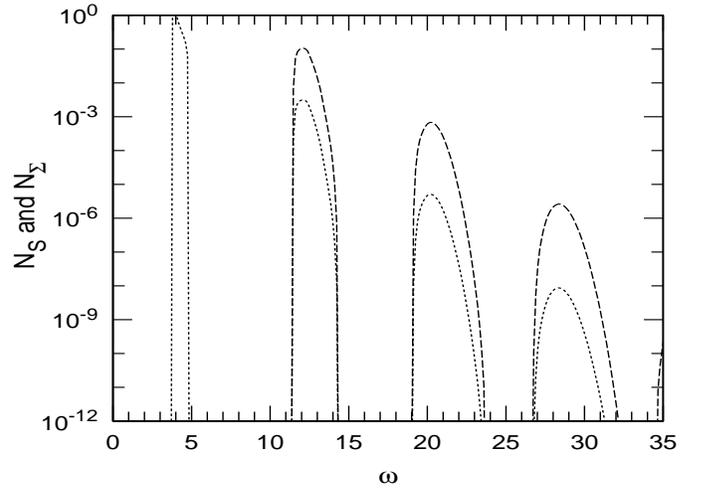}}
\end{picture}
\caption{Density of states $N_{\scriptstyle \rm S}$ 
	(short dashed curve)
	and spectral function $N_\Sigma$ (long dashed curve)
	in the AB-CDW at $U=8$
	in logarithmic scale. The difference quantities are not shown
	since their values lie only slightly under those of the sum
	quantities.}
	\label{fi:9}
\end{figure}

In fig.\ \ref{fi:8}, hardly discernible
satellite bands exist at $\omega\approx12$. They are
engendered by the finite imaginary part of the self-energy
at these energies (see fig.\ \ref{fi:8}(b) and (d)).
To demonstrate that there are in fact infinitely many
satellite bands
with exponentially decreasing weights, the densities
 $N_{\scriptstyle\rm S}$ and $N_\Sigma$ are
plotted logarithmically in fig.\ \ref{fi:9}.
The
principal band of the DOS consists of quasi-particles with infinite
life-time at $\omega_\Delta\approx 4$. The satellite bands
correspond to peaks in the spectral density of the self-energy.
The satellite bands are located at $(2m+1)\omega_\Delta$ where $m$ is
an integer. The peaks in the spectral density of the self-energy
are located at $(2m+1)\omega_\Delta$ where $m$ is
an integer but {\em not} $0$ or $-1$. This phenomenon is
generic for the self-consistent solution of a system of
equation comprising convolutions of strongly peaked
functions. It appears only at large values of $U$
because it is necessary that $\omega_\Delta\approx U/2$ is larger
than the band width in order to resolve the peaks.
Note that according to  (\ref{green3}), a large value
of $\Delta$ induces band narrowing. Whereas the principal
band is $\sqrt{6}$ wide at $U=0$, its width is shrunk to about unity
 in fig.\ \ref{fi:8}(a).

For detailed numerical results on the order parameter
as function of interaction and of temperature as well as on
the critical temperature the reader is referred to
ref.\ 6. The asymptotic behavior at small
$U$ is discussed analytically by van Dongen
\cite{donge91,donge94b}. In a nutshell, the correlation term
renormalizes the Hartree results for $b$ and $T_{\scriptstyle\rm c}$
by a constant factor of order unity which tends to unity
for $d\to \infty$.

\section{Conductivity: Foundations}
Due to the point symmetry group of the hypercubic
lattices the conductivity $\sigma(\omega)$ can be
treated as a scalar. Previous one-particle results
showed that the treatment on the level of linear
$1/d$ corrections should yield reasonable results \cite{halvo94}
in $d=3$.

\begin{figure}[hbt]
\setlength{\unitlength}{1cm}
\begin{picture}(8.2,5.8)(0,0)
\put(1,0.6){\psfig{file=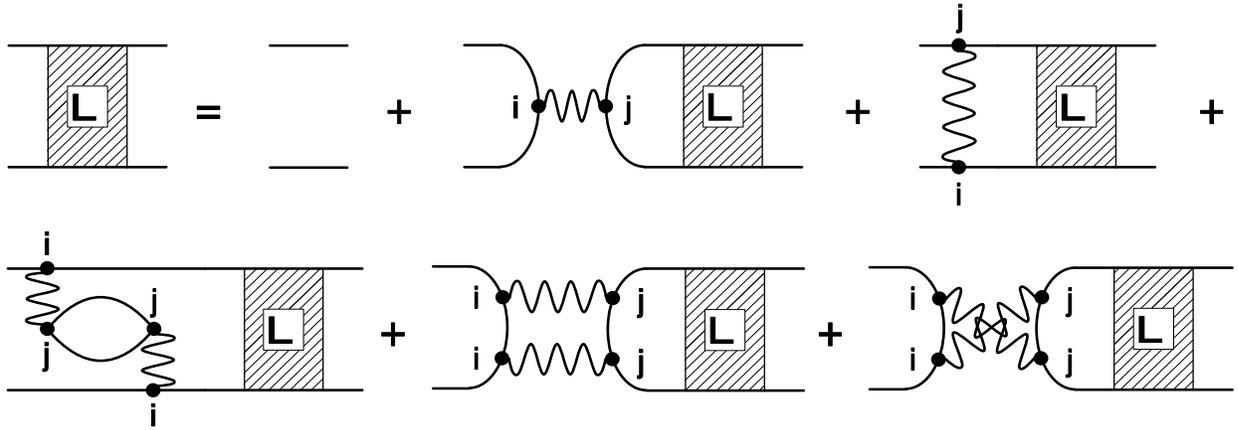,height=5.8cm,width=16.4cm,angle=270}}
\end{picture}
\caption{Diagrammatic representation 
	of the Bethe-Salpeter equation
	resulting from $\Phi_{\scriptstyle \rm A}$
	according to Baym/Kadanoff.
	The wavy lines stand for interaction; the solid ones for
	fermionic propagators. The direction of the lower propagators
	is opposite to the one of the upper propagators.}
	\label{fi:19}
\end{figure}

The conductivity
is calculated from a two-particle correlation function.
This will be done here from the current-current
correlation function $\chi^{\scriptstyle\rm JJ}$. The conductivity
comprises two contributions
$\sigma(\omega) = \sigma_1(\omega) +\sigma_2(\omega)$.
The first term depends on the  occupation of the
momentum states $\langle {\hat n}_{\bf k}\rangle$
whereas the second term is proportional to
$\chi^{\scriptstyle\rm JJ}(\omega)$ \cite{mahan90}
\begin{mathletters}\label{a2.6}
\begin{eqnarray}\label{a2.6a}
\sigma_1(\omega) &=& \frac{i}{\omega} \int\limits_{\scriptstyle\rm BZ}
\frac{\partial^2\varepsilon({\bf k})}{\partial k_1^2}
 \langle {\hat n}_{\bf k}\rangle
\frac{dk^d}{(2\pi)^d}
\\ \label{a2.6b}
\sigma_2(\omega) &=& \frac{i}{\omega}\chi^{\scriptstyle\rm JJ}(\omega)
\ .\end{eqnarray}
\end{mathletters}
The current-currrent correlation function will be
computed including $1/d$ corrections with the help
of the Baym/Kadanoff formalism \cite{baym61,baym62}.
Specific correlation functions are determined from
the general two-particle correlation function
$L(12,1'2')$ via
\begin{equation}\label{baym0}
\chi^{AB} = \int A(1,1')L(12,1'2')B(2,2')d11'22'\ .
\end{equation}
The numbers stand for composite space and time
coordinates (or momentum and frequency coordinates).
The measure $d11'22'$ tells which coordinates are
integrated. The quantities $A$ and $B$ represent the
operators for which the correlation function is computed.
The Bethe-Salpeter equation determines $L(12,1'2')$
implicitly using the kernel (or effective two-particle
interaction) $\Xi(35,46)$ and the Green function
$G(1,2)$
\begin{equation}\label{baym1}
L(12,1'2') = G(1,2')G(2,1') +\int
G(1,3) G(1',4) \Xi(35,46) L(62,52') d3456 \ .
\end{equation}
Like the kernel of the Dyson equation, namely the
self-energy, the kernel $\Xi(35,46)$ of the
Bethe-Salpeter equation is given as functional derivative
with respect to the Green function
\begin{equation}\label{funcder2}
\Xi(35,46) = \frac{\partial\Sigma(3,4)}{\partial G(6,5)}
= \frac{\partial^2\Phi}{\partial G(4,3)\partial G(6,5)}
\ .\end{equation}
Diagrammatically, the functional derivation is the omission
of a propagator line. Applying these steps to the
approximate generating functional $\Phi_{\scriptstyle\rm A}$ in
fig.\ \ref{fi:3} yields the diagrammatic representation
of the Bethe-Salpeter equation (\ref{baym1}) in fig.\
\ref{fi:19}. 
The first diagram with a wavy interaction
line in the upper row  stems from the Hartree diagram,
the last diagram in the upper row results from the Fock
diagram. The diagrams in the lower row in fig.\ \ref{fi:19}
are generated by the different possibilities to take out
two propagator lines from the correlation diagram.

\begin{figure}[hbt]
\setlength{\unitlength}{1cm}
\begin{picture}(8.2,1.7)(0,0)
\put(9,0.6){\psfig{file=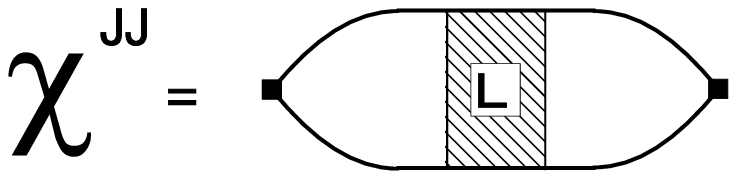,height=1.7cm,width=7.4cm}}
\end{picture}
\caption{Diagrammatic representation 
	of eq.\ (\protect\ref{baym0}) for the
	current-currrent correlation.}
	\label{fi:20}
\end{figure}

\begin{figure}[hbt]
\setlength{\unitlength}{1cm}
\begin{picture}(8.2,1.6)(0,0)
\put(0.8,0.6){\psfig{file=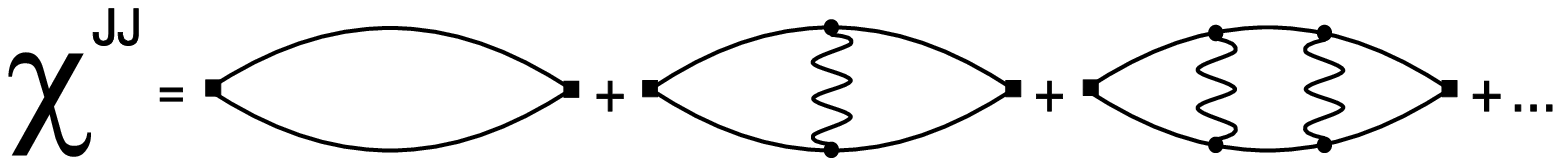,height=1.6cm,width=15.7cm}}
\end{picture}
\caption{Current-currrent correlation with $1/d$ corrections.}
\label{fi:21}
\end{figure}

Fortunately, the summation  in fig.\ \ref{fi:19}
simplifies considerably for the evaluation of the
current-current correlation function $\chi^{\scriptstyle\rm JJ}$. Fig.\
\ref{fi:20}
 displays equation (\ref{baym0}). The squares
represent the current vertices
\begin{equation}\label{strver0}
\mbox{J}(1,1') = \delta({\bf k}_1-{\bf k}_{1'})
\delta(\omega_1-\omega_{1'}-\omega)
\frac{\partial \varepsilon}{\partial k_{1,1}}
\ .\end{equation}
Due to symmetry it does not matter for which spatial direction
$\mbox{J}(1,1')$ is calculated; $k_{1,1}$ is one
arbitrarily chosen component.
The crucial property of the current vertex is its oddness
as function of $k_{1,1}$. All interaction terms
which are even in $k_{1,1}$ do not contribute. This is the
case for all the diagrams resulting from the local
correlation in the lower row and for the diagram coming
from the Hartree term since only {\em one} site appears
on either side. Hence, only the geometric series
depicted in fig.\ \ref{fi:21}
caused by the non local Fock
term is left.
For comparison: in the infinite dimensional Hubbard model the
simplifications are even more drastic. All vertex corrections
drop out and the current-current correlation function is just the
convolution of two Green functions \cite{khura90}.

Let us call the value of the first diagram in fig.\ \ref{fi:21}
the ``free'' current-current correlation function and let us use the
symbol $\chi^{\scriptstyle\rm JJ}_0$ for it.
In the homogeneous phase  one obtains with the help of
(\ref{baym0}), (\ref{strver0}) and of the propagator in ${\bf k}$-space
$(\omega-\Sigma(\omega)-\gamma\varepsilon({\bf k}))^{-1}$
\begin{equation}\label{freihom}
\chi^{\scriptstyle\rm JJ}_0(i\omega_m) =
\frac{4T}{Z} \sum\limits_{\omega_\nu-\omega_\lambda=\omega_m}
 \;\int\limits_{\scriptstyle\rm BZ} \frac{\sin^2(k_1)}
{\left(w_{\nu}-\gamma\varepsilon({\bf k})\right)
\left(w_\lambda-\gamma\varepsilon({\bf k})\right)}
\frac{dk^d}{(2\pi)^d}
\ ,\qquad\quad\end{equation}
where $w_{\nu/\lambda}:=i\omega_{\nu/\lambda}-
\Sigma(i\omega_{\nu/\lambda})$.

We focus now on the segments between two wavy lines in fig.\ \ref{fi:21}.
The conservation of energy and of momentum makes
it possible to carry out the sum over all momentums and energies
by considering independent momentums and energies circulating in
each segment. Then the momentum {\em in} a wavy line is the difference
of two adjacent wave vectors ${\bf k}$ and
${\bf k'}$. A second time,
the evenness and the oddness in the components of the wave vector is used
to write for the factor of an interaction line
\begin{eqnarray}
-\frac{U}{d}\sum\limits_{i=1}^d\cos(k_i- k_i')&=&
-\frac{2U}{Z}\sum\limits_{i=1}^d
\left[\sin(k_i)\sin(k_i')+\cos(k_i)\cos(k_i')\right]
\nonumber \\
&\to& -\frac{2U}{Z} \sin(k_1)\sin(k_1')
\ .
\label{well}
\end{eqnarray}
The argument is obvious for one of the border segments and follows
for those in the middle by induction.

At the end one realizes that each segment corresponds to a
factor of $-(U/2)\chi^{\scriptstyle\rm JJ}_0$ which justifies to call
the right side of fig.\ \ref{fi:21} a geometric series
which takes the value
\begin{equation}\label{vollchi}
\chi^{\scriptstyle\rm JJ}(\omega+0i) =
 \frac{\chi^{\scriptstyle\rm JJ}_0
(\omega+0i)}{1+U\chi^{\scriptstyle\rm JJ}_0(\omega+0i)/2}
\end{equation}
after analytic continuation. The derivation of a similar
formulae in the AB-CDW is given in appendix B. The results are
cited below.

The momentum integration in (\ref{freihom}) requires a modified
DOS, to be called the conductivity DOS henceforth
\begin{equation}\label{ldich0}
N_{c,0}(\omega) :=
\int\limits_{\scriptstyle\rm BZ}\sin^2(k_1)
\delta(\omega-\varepsilon({\bf k}))\frac{dk^d}{(2\pi)^d}
\ ,\end{equation}
from which we define also the conductivity Green function
$g_{c,0}(z) := \int\limits_{-\infty}^\infty
N_{c,0}(\omega)/(z-\omega) d\omega$.
The conductivity DOS can be simply derived once the DOS
is known. These two functions are related via
\begin{equation}\label{zufo3}
N_0(\omega) = -\frac{2}{\omega}
\frac{\partial N_{c,0}}{\partial\omega}(\omega) \ .
\end{equation}
This relation stems from the fact that one has to replace
one of the $d$ factors $(1/\pi) 1/\sqrt{t^2-\omega^2}$ in the convolution
for the DOS by $(1/\pi) \sqrt{t^2-\omega^2}$ in order to calculate the
conductivity DOS. The derivation uses the representation of
convolutions as products in Fourier space.

Using the definition of the conductivity Green function and
partial fraction expansion it is straightforward to rewrite
(\ref{freihom})
\begin{equation}\label{hom0}
\chi^{\scriptstyle\rm JJ}_0(i\omega_m) = - \frac{4T}{\gamma Z}
\sum\limits_{\omega_\nu-\omega_\lambda=\omega_m}
\frac{g_c(i\omega_\nu/\gamma)-g_c(i\omega_\lambda/\gamma)}
{w_\nu-w_\lambda} \ .
\end{equation}
Analytic continuation of the latter gives the general formula
(eq.\ (14) in ref.\ 26) 
for the current-current correlation function in the homogeneous phase.

In the AB-CDW, it is also possible to sum the series in
fig.\ \ref{fi:21} as geometric series. The main difference is
the fact that $2\times2$ matrices instead of scalars are
involved. The details are given in appendix B; the results
\cite{uhrig95c} are
\begin{equation}\label{vollchi2}
\chi^{\scriptstyle\rm JJ}(i\omega_m) = \frac{2}{U}-\frac{2}{U}\frac{1-A_2}{
(1-A_1)(1-A_2)-A^2_3}
\ ,\end{equation}
where the quantities $A_1, A_2$ and $A_3$ are defined by
\begin{mathletters}\label{basis0}
\begin{eqnarray}
A_1(i\omega_m) &=&
\frac{2UT}{Z}\sum\limits_{\omega_\nu-\omega_\lambda=\omega_m}
\nonumber \\
&&\hspace*{-0.8cm} \left[
\frac{(w_\lambda+w_\nu)(g_{c,\scriptstyle\rm S}(i\omega_\nu)-
g_{c,\scriptstyle\rm S}(i\omega_\lambda))-
(\Delta(i\omega_\lambda)+\Delta(i\omega_\nu))
(g_{c,\scriptstyle\rm D}(i\omega_\nu)-g_{c,\scriptstyle\rm D}
(i\omega_\lambda))}
{w_\nu^2-w_\lambda^2-(\Delta^2(i\omega_\nu)-\Delta^2(i\omega_\lambda))}
\right]
\label{basis0a} \\
A_2(i\omega_m) &=& 
\frac{2UT}{Z}\sum\limits_{\omega_\nu-\omega_\lambda=\omega_m}
\nonumber \\
&&\hspace*{-0.8cm} \left[
\frac{(w_\lambda-w_\nu)(g_{c,\scriptstyle\rm S}(i\omega_\nu)+
g_{c,\scriptstyle\rm S}(i\omega_\lambda))-
(\Delta(i\omega_\lambda)-\Delta(i\omega_\nu))
(g_{c,\scriptstyle\rm D}(i\omega_\nu)+g_{c,\scriptstyle\rm D}
(i\omega_\lambda))}
{w_\nu^2-w_\lambda^2-(\Delta^2(i\omega_\nu)-\Delta^2(i\omega_\lambda))}
\right]
\label{basis0b} \\
A_3(i\omega_m) &=&
\frac{2UT}{Z}\sum\limits_{\omega_\nu-\omega_\lambda=\omega_m}
\nonumber \\
&&\hspace*{-0.8cm} \left[
\frac{\Delta(i\omega_\lambda)g_{c,\scriptstyle\rm S}(i\omega_\nu)-
w_\lambda g_{c,\scriptstyle\rm D}(i\omega_\nu)+
\Delta(i\omega_\nu)g_{c,\scriptstyle\rm S}(i\omega_\lambda)-
w_\nu g_{c,\scriptstyle\rm D}(i\omega_\lambda)}
{w_\nu^2-w_\lambda^2-(\Delta^2(i\omega_\nu)-\Delta^2(i\omega_\lambda))}
\right]\ .   \label{basis0c}
\end{eqnarray}
\end{mathletters}
In complete analogy to the usual Green functions, the
conductivity Green functions are
$g_{c,\scriptstyle\rm S} := (g_{c,\scriptstyle\rm A}+
g_{c,\scriptstyle\rm B})/2$ and
$g_{c,\scriptstyle\rm D} := (g_{c,\scriptstyle\rm A}-
g_{c,\scriptstyle\rm B})/2$, hence
\begin{mathletters}\label{lgreen3}
\begin{eqnarray} \label{lgreen3a}
g_{c,\scriptstyle\rm S}(\omega)&=&
\frac{\omega-\Sigma(\omega)}{\gamma\sqrt{\left(\omega-
\Sigma(\omega)\right)^2 -\Delta^2(\omega)}}
g_{c,0}(\sqrt{w^2-\Delta^2(\omega)}/\gamma)
\\  \label{lgreen3b}
g_{c,\scriptstyle\rm D}(\omega)&=&
\frac{\Delta(\omega)}{\gamma
\sqrt{w^2-\Delta^2(\omega)}}
g_{c,0}(\sqrt{w^2-\Delta^2(\omega)}/\gamma)
\ ,
\end{eqnarray}
\end{mathletters}
which compares to (\ref{green3})
 ($w$  is short-hand for $\omega-\Sigma(\omega)$).

Now a relation for the dc-conductivity shall be derived.
In order that the limit
$\lim_{\omega\to0} \sigma(\omega)$ exists
\begin{equation}\label{wfsum0}
\chi^{\scriptstyle\rm JJ}(0)  =
\int\limits_{\scriptstyle\rm BZ}
\frac{\partial^2\varepsilon}{\partial k_1^2}({\bf k})
\langle {\hat n}_{\bf k}\rangle \frac{dk^d}{(2\pi)^d}
 \; =\; \frac{\langle \hat{T}\rangle}{d}
\end{equation}
must hold according to (\ref{a2.6}). The operator $\hat {T}$
stands for the kinetic energy.
Eq.\ (\ref{wfsum0}) implies also the $f$-sum rule
$\int\limits_{-\infty}^\infty (i\chi^{\scriptstyle\rm JJ}/\omega) d\omega =
 -\pi \langle \hat{T}\rangle/d$. At the end of
 appendix B, it is shown explicitly
that (\ref{wfsum0}) is valid since $A_3$ vanishes at $\omega=0$
and $A_1=-U\langle \hat{T}\rangle/(2 \gamma d)=1-1/\gamma$.
For the dc-conductivity one obtains
\begin{equation}\label{dcltf}
\sigma(0) = i \left.\frac{\partial\chi^{\scriptstyle\rm JJ}}{\partial\omega}
\right|_{\omega=0}
= -\frac{2i \gamma^2}{U}
\left.\frac{\partial A_1}{\partial\omega}\right|_{\omega=0}
\ .\end{equation}
For explicit evaluation it is useful to split 
$\sigma_{\scriptstyle\rm dc}(0)$
into a term including retarded and advanced Green functions
$\sigma_{\scriptstyle\rm dc1}$ and
a term including only retarded or advanced Green functions
$\sigma_{\scriptstyle\rm dc2}$ after analytic continuation. This yields
\begin{equation}\label{dcltf1}
\sigma_{\scriptstyle\rm dc1} = \frac{(\gamma)^2}{\pi Z}
\int\limits_{-\infty}^{\infty} \frac{(1-{\mathop{\rm Re}\nolimits}\Sigma)
N_{c,\scriptstyle\rm S}-({\mathop{\rm
Re}\nolimits} \Delta) N_{c,\scriptstyle\rm D}}
{(1-{\mathop{\rm Re}\nolimits}\Sigma)N_\Sigma+
({\mathop{\rm Re}\nolimits} \Delta) N_\Delta}
(-f_{\scriptstyle\rm F}'(\omega)) d\omega
\ ,\end{equation}
where $f_{\scriptstyle\rm F}'(\omega)$  is 
the derivative of the Fermi distribution, and
\begin{eqnarray}
\sigma_{\scriptstyle\rm dc2} &=& -\frac{\gamma^2}{\pi Z}
\int\limits_{-\infty}^{\infty} \left.\frac{(1-\Sigma)
\partial_\omega g_{c,\scriptstyle\rm S}-
\Delta \partial_\omega g_{c,\scriptstyle\rm D}}
{(1-\Sigma)(\partial_\omega\Sigma-1)+\Delta
 \partial_\omega\Delta}\right|_{\omega+0i}
 (-f_{\scriptstyle\rm F}'(\omega)) d\omega
\nonumber \\ \label{dcltf2}
&=& \frac{1}{\pi Z}
\left[1-{\mathop{\rm Re}\nolimits} \int\limits_{-\infty}^{\infty}
\left((\omega-\Sigma)g_{\scriptstyle\rm S}-
\Delta g_{\scriptstyle\rm D}\right)_{\omega+0i}
  (-f_{\scriptstyle\rm F}'(\omega)) d\omega \right]
\ .
\end{eqnarray}
In the last expressions, all the Green functions are retarded.
In the homogeneous phase, the contribution (\ref{dcltf1})
is more important than the one in (\ref{dcltf2}). The former diverges
for $T\to 0$ and $\omega\to 0$, the latter does not. In the symmetry
broken AB-CDW, however, both terms turn out to be essential.

Eqs.\ (\ref{vollchi2}), (\ref{basis0}), (\ref{dcltf1}) and (\ref{dcltf2})
are the foundation for the calculation of the conductivity for
zero and for non-zero order parameter. The focus of the present
work is on the AB-CDW. The properties of the conductivity
in the homogeneous phase (e.g.\ Fermi liquid behavior) are presented
in detail in ref.\ 26 
 where also the influence of the
truncation of the $1/d$ expansion is discussed.

\section{Conductivity: Results}
In this section we present and discuss results which follow
from the general equations derived in the previous section.
All results are calculated at half-filling and for
$d=3$.

\begin{figure}[hbt]
\setlength{\unitlength}{1cm}
\begin{picture}(8.2,7)(0,0.7)
\put(8.3,0){\psfig{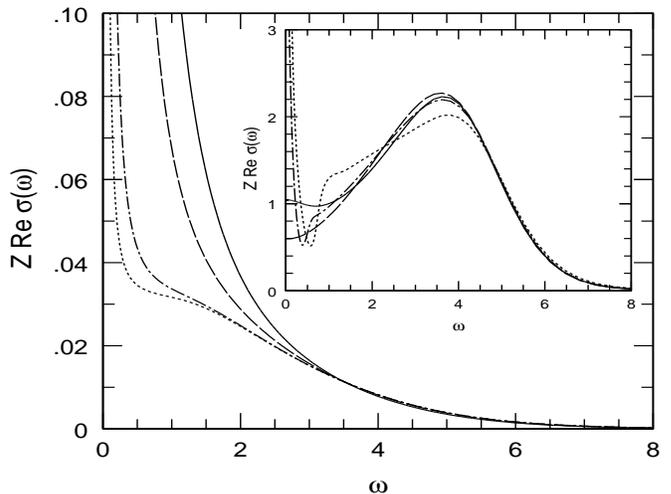}}
\end{picture}
\caption{Scaled real part of the dynamic 
	conductivity Re$\ \sigma(\omega)$ in the non symmetry broken
	phase at $U=4.243$ for $T=0.393$ (solid lines),
	$T=0.196$ (dashed lines), $T=0.049$ (dashed-dotted lines),
	and $T=0.025$ (dotted lines). Main figure: 
	spinless fermions in $d=3$; inset: Hubbard model in the
	non-crossing approximation
	(data from Th.~Pruschke).}
	\label{fi:res1}
\end{figure}
In fig.\ \ref{fi:res1}, the real part of the
dynamic conductivity is depicted in the non symmetry broken
phase for different temperatures, i.e.\ the occurrence of
a symmetry broken phase at low temperature is discarded
deliberately for the moment.
 They are compared with results of Pruschke, Cox, and Jarrell
\cite{prusc93a,prusc93b}
for the half-filled Hubbard model in $d=\infty$, obtained in the
 non-crossing approximation. In both cases the interaction
value is $U=4.243$ (in our units)
 which is just below the value where the
Mott-Hubbard transition occurs in the Hubbard model \cite{prusc93b}.
For spinless fermions the Drude peak is absolutely dominant.
Its weight is very large. Its width is given by the
imaginary part of the self-energy at the Fermi level
$N_\Sigma(0)$ (see (\ref{basis0}) with $\Delta=0$ or
eq.\ (14) in ref.\ 26) 
, i.e.\ the width
is proportional to $T^2$. The shape of the Drude peak
corresponds very well to a lorenzian.

Only at low temperatures a
shoulder emerges. This shoulder is the effect of
interaction induced scattering. The fluctuations are not
particularly strong. It was already shown previously
\cite{uhrig95a} that the average over the $Z$ interaction
partners reduces the relative fluctuations. There is no Mott-Hubbard
transition without symmetry breaking in the spinless fermion model
because an increasing interaction enhances not only the
fluctuations but also the Fock term (absent in the Hubbard
model) which stabilizes the Fermi liquid phase.
These features are particularly obvious in the comparison with
the Hubbard model data. In this model, the Drude peak
is very reduced at all displayed temperatures since much of the
weight is shifted to the peaks induced by the strong
local particle density fluctuations.

Besides the difference
shoulder vs.\ peak it is interesting to note the difference in
energy scales. In the Hubbard model, it is more or less
$U$ which sets the energy at which the peak occurs. This can be
understood as the energetic effect of whether or not an electron
with a different spin is present.
The typical energy for the shoulder is obviously much smaller.
This in turn can be understood in the same way as before but it
has to be taken into account that the number of possible interaction
partners $Z$ leads to a reduction of the relative
 fluctuations of the order of
$1/\sqrt{Z}$. This yields  an energy
of roughly $1.7$ in the particular example 
which is in good agreement with the numerical result.

\begin{figure}[hbt]
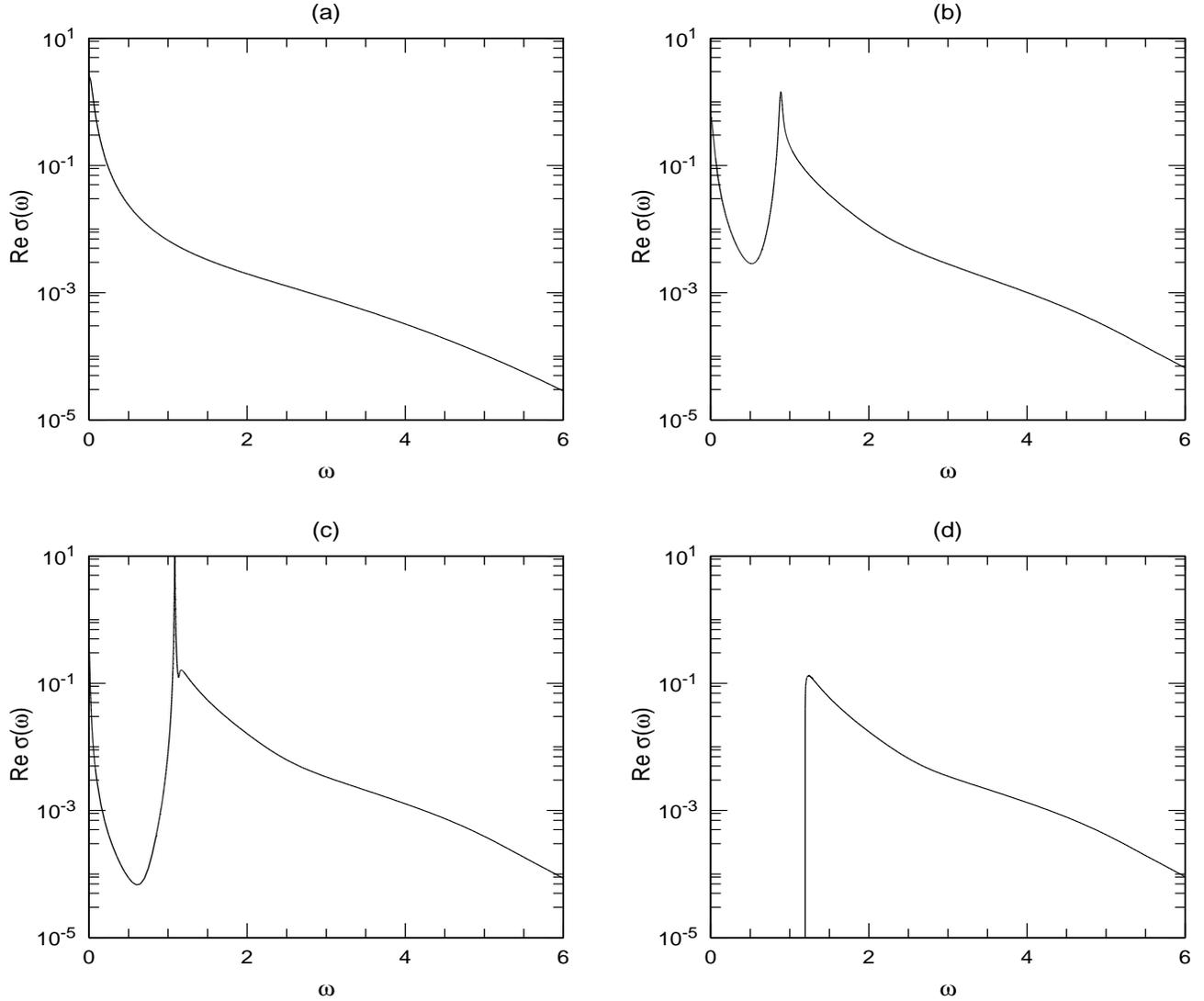

\setlength{\unitlength}{1cm}
\begin{picture}(8.2,15)(0,0.7)
\put(-0.7,7.5){\psfig{file=fig15a.ps,height=7.2cm,width=8cm,angle=270}}
\put(8.3,7.5){\psfig{file=fig15b.ps,height=7.2cm,width=8cm,angle=270}}
\put(-0.7,0){\psfig{file=fig15c.ps,height=7.2cm,width=8cm,angle=270}}
\put(8.3,0){\psfig{file=fig15d.ps,height=7.2cm,width=8cm,angle=270}}
\end{picture}
\caption{Real part of the dynamic 
	conductivity Re$\ \sigma(\omega)$
	in $d=3$ at $U=2.0$ in logarithmic scale. Fig.\ (a)
	$T=0.300000$ and $b=0$; fig.\ (b) $T=0.225658$ and $b=0.250000$;
	fig.\ (c) $T=0.155286$ and $b=0.299801$; fig.\ (d) 
	$T=0$ and $b=0.311005$. In fig.\ (d) the $\delta$-peak
	at $\omega=0.13312$ is not shown, its weight is 0.062336.}
	\label{fi:23}
\end{figure}

Due to the nesting at half-filling, the system of spinless fermions
undergoes a transition to a spontaneously broken translation
symmetry for all (positive) values of the interaction
on lowering the temperature. This spontaneously broken
discrete symmetry implies the occurrence of a gap which
grows exponentially $\omega_\Delta \propto \exp(-c/U)$
for low values of the interaction at $T=0$ (see ref.\ 6 
and refs.\ therein).
It is visible in the dynamic conductivity\cite{uhrig95c}.
 In fig.\ \ref{fi:23},
its growth on decreasing temperature is shown in four snap-shots.
In fig.\ \ref{fi:23}(a), $T$ is still above its critical value. No
structure is visible except for the dominant Drude peak already
discussed in fig.\ \ref{fi:res1}.
In figs.\ \ref{fi:23}(b)-(d) the gap is present and
discernible. Its value is
approximately $2\omega_\Delta$ if $\omega_\Delta$ is the value of the
energy gap in the DOS, see figs.\ \ref{fi:7} and \ref{fi:8}.
 But there is also
some weight within the gap for $T>0$ since the correlation
contribution blurred already the gap in the DOS.
Note in passing that the f-sum rule can be verified numerically on
the results shown in fig.\ \ref{fi:23} very accurately
(to the fraction of a percent at $T=0$;
to the fraction of a permille in the homogeneous phase).

The Drude peak does not vanish immediately in the AB-CDW.
It becomes smaller and narrower on decreasing temperature.
Its maximum value does not vanish for $T\to 0$ (see below) but
its weight does. In fig.\ \ref{fi:24},
\begin{figure}[hbt]
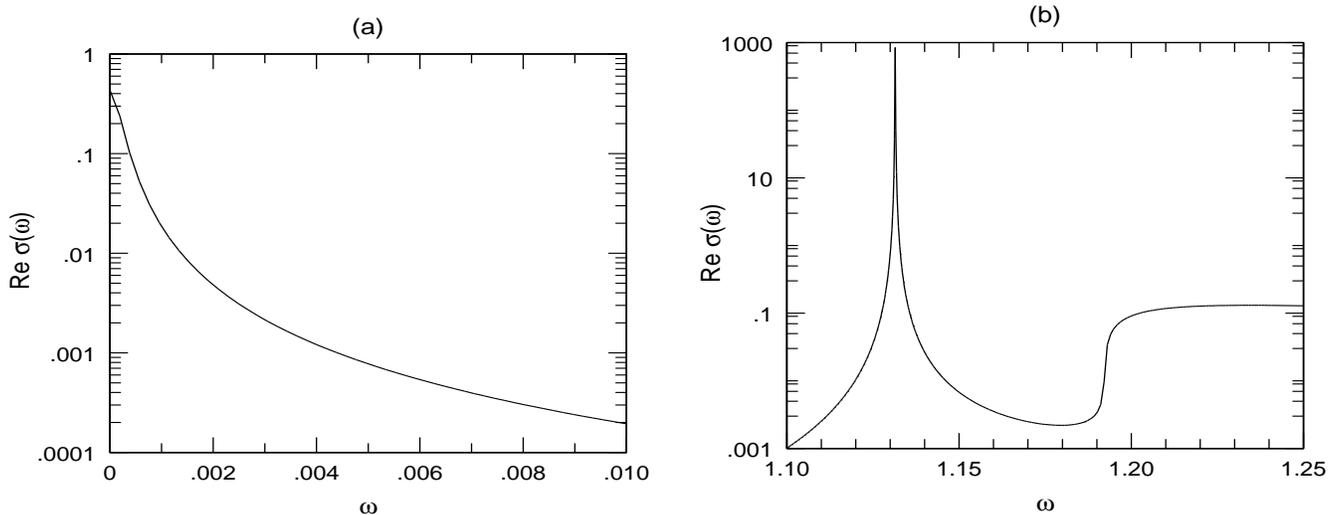

\setlength{\unitlength}{1cm}
\begin{picture}(8.2,7)(0,0.7)
\put(-0.7,0){\psfig{file=fig16a.ps,height=7.2cm,width=8cm,angle=270}}
\put(8.3,0){\psfig{file=fig16b.ps,height=7.2cm,width=8cm,angle=270}}
\end{picture}
\caption{Enlargements
	of two frequency intervals for $T=0.0833833$ and
	$b=0.310773$. Fig.\ (a) shows details of the Drude peak;
	fig.\ (b) the excitonic resonance.}
	\label{fi:24}
\end{figure}
two frequency intervals
are shown in detail for a fairly low temperature. Fig.\ \ref{fi:24}(a)
displays the Drude peak again. The interesting feature 
 is its small width (compared with the width of the Drude
peaks in figs.\ \ref{fi:23}(b) and (c)). It cannot be explained
by a factor of $T^2$ but corresponds to an exponential shrinking
$\exp(-\omega_\Delta/T)$. As already observed  in the
one-particle properties, an increasing gap reduces the influence
of the fluctuations.

Fig.\ \ref{fi:24}(b) shows a very interesting feature below the
proper band edge at $\omega\approx 2\omega_\Delta$. This  resonance is also
visible in fig.\ \ref{fi:23}(c) whereas the resonance and the
band edge are not resolved at a higher temperature, fig.\
\ref{fi:23}(b). The resonance can very well be approximated by
a lorenzian. At $T=0$, it is also present as a $\delta$-peak
(not shown in fig.\ \ref{fi:23}(d)). It originates from
a zero of the denominator in (\ref{vollchi2}). At $T>0$, only
the real part of the denominator vanishes and its imaginary
part leads to the observed broadening which depends
strongly, namely exponentially, on the temperature.

Physically the resonance can be interpreted as a bound state,
an exciton, between a particle in the upper band and a hole
in the lower band in the reduced Brillouin zone of the AB-CDW.
The energy difference between the position of the exciton and the
band edge is its binding energy. The type of diagrams which
yield the denominator in (\ref{vollchi2}) corroborates the interpretation
as an exciton. The vertical interaction lines stand for the
repeated interaction between particle and hole in the two
propagators involved in the calculation of $\chi^{\scriptstyle\rm JJ}$.
It should be noted that, for instance, for the parameters of
fig.\ \ref{fi:23}(d) about 70\% of the weight of the conductivity
are found in the excitonic resonance (one may not be misled
by the logarithmic scale). This means that the excitonic
effect is not at all a small side effect.

\begin{figure}[hbt]
\setlength{\unitlength}{1cm}
\begin{picture}(8.2,7)(0,0.7)
\put(8.3,0){\psfig{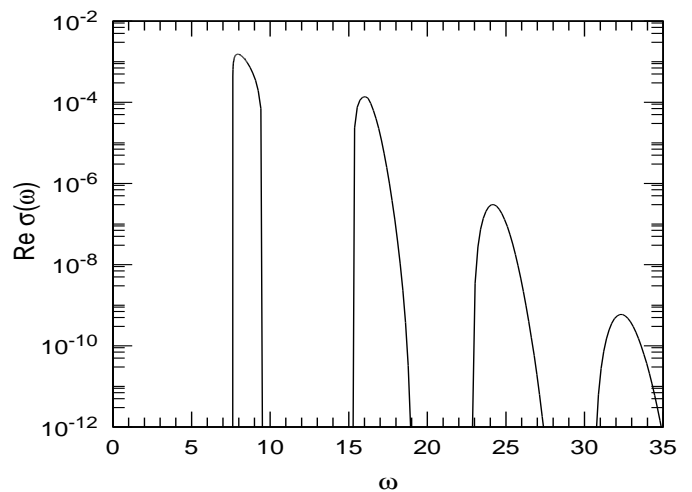}}
\end{picture}
\caption{Real part of the dynamic conductivity Re$\ 
	\sigma(\omega)$
	in $d=3$ at $U=8.0$ for $T=0$ in logarithmic scale. The
	$\delta$-distribution is not displayed.}
	\label{fi:25}
\end{figure}

Concluding the part on the dynamic conductivity, we
discuss fig.\ \ref{fi:25}
which shows results for
a large interaction value $U$. Due to the induced large gap
and due to the narrow
effective band width several
frequency intervals of absorption are well separated.
The peaks are caused by the convolution of the satellite
band presented for the one-particle properties.
Note, however, that the weight of these satellites
decreases rapidly by a factor of 100 from peak to peak.
These small amplitudes render an experimental verification
certainly extremely difficult if not impossible.
Nevertheless, it would be interesting to know whether such satellites
exist. Their existence would support the application
of a self-consistent approximation since the non self-consistent
calculation yields only two peaks besides the $\delta$-peak
which is not shown.

\begin{figure}[hbt]
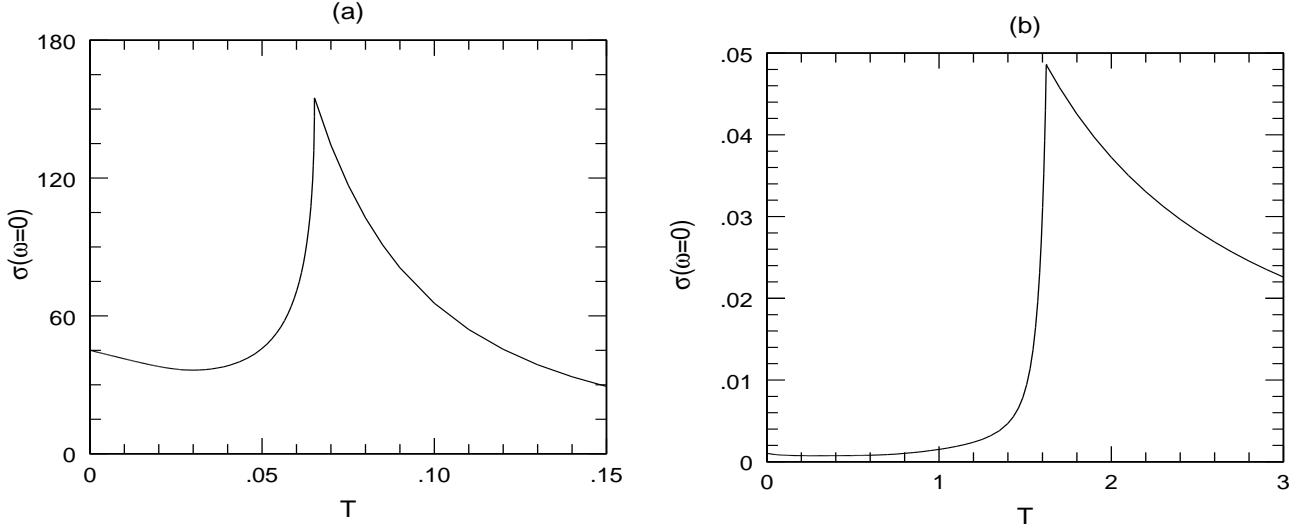

\setlength{\unitlength}{1cm}
\begin{picture}(8.2,7)(0,0.7)
\put(-0.7,0){\psfig{file=fig18a.ps,height=7.2cm,width=8cm,angle=270}}
\put(8.3,0){\psfig{file=fig18b.ps,height=7.2cm,width=8cm,angle=270}}
\end{picture}
\caption{Temperature dependence of the dc-conductivity 
	at $U=1.0$ (fig.\  (a))
	and at $U=8.0$ (fig.\ (b)). Below $T=0.026$ in fig.\ (a) and below
	$T=0.6$ in fig.\ (b) a fit was used (see main text).}
	\label{fi:26}
\end{figure}

Since the dc-conductivity in absence of symmetry breaking
has been extensively discussed
in ref.\ 26 
we will treat here exclusively the case with symmetry breaking.
The result of (\ref{dcltf1}) and (\ref{dcltf2}) is depicted
in fig.\ \ref{fi:26}
for weak and strong interaction \cite{uhrig95c}.
To the right of the cusp the system is in the
non symmetry broken phase. The conductivity is
essentially proportional \cite{uhrig95a} to $T^2$.
On entering the symmetry broken phase with gap, the
conductivity falls drastically since the energy gap
reduces the DOS at the Fermi level.
Surprisingly, however, the conductivity does {\em not} vanish
for $T\to 0$ although the DOS vanishes in this limit.
There is even a very slight uprise of $\sigma_{\scriptstyle\rm dc}$ close
to $T=0$.
This phenomenon is again a manifestation of the suppression
of correlation effects by the energy gap. The DOS is
reduced by a factor of $\exp(-\omega_\Delta/T)$ but so is the
imaginary part of the self-energy in (\ref{basis0}) which
is responsible for the quasi-particle life-time.
These two effects cancel exactly.
Put differently, an exponentially small number of quasi-particles
of exponentially large life-time
carries a constant current (but see discussion below).
It remains an algebraic dependence on $T$
of the dc-conductivity. The constant term and the
linear one can be computed analytically and where
used to complete the curves in fig.\ \ref{fi:26} for
small values of $T$ where the numerical calculation
is no longer precise enough due to extinction.

\begin{figure}[hbt]
\setlength{\unitlength}{1cm}
\begin{picture}(8.2,7)(0,0.7)
\put(8.3,0){\psfig{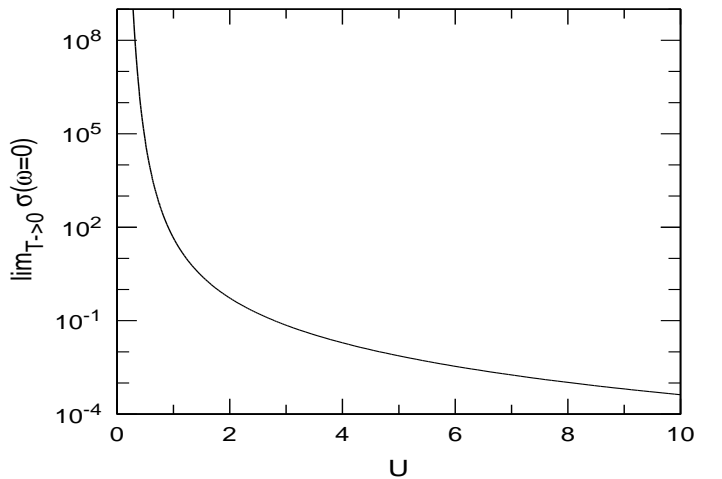}}
\end{picture}
\caption{dc-conductivity $\sigma(\omega=0)$ in the
	 limit $T\to 0$ in logarithmic scale.}
	\label{fi:27}
\end{figure}

The limit value $\lim_{T\to 0} \sigma(\omega=0)$ is given
in fig.\ \ref{fi:27}
as function of $U$. As expected
it decreases rapidly for $U\to \infty$. Note the logarithmic
scale.

What do the above findings for $\sigma_{\scriptstyle\rm dc}$ imply
for the existence of a metal-insulator transition?
Seemingly, even spontaneous symmetry breaking
does not suffice to render the system insulating.
But it must be noted that the ``residual''
conductivity $\lim_{T\to 0} \sigma(\omega=0)$
is infinitely fragile: any other
arbitrarily weak scattering mechanism
which does not die out on $T\to 0$
e.g.\ disorder or scattering at the borders of the sample, will take
over. The exponentially vanishing DOS will yield an
exponentially vanishing dc-conductivity.
This is reflected in the exponentially decreasing
width of the Drude peak which, at constant height,
implies an exponentially decreasing weight.

Experimentally, very pure samples might allow to see
the beginning of the plateaus in fig.\ \ref{fi:26}
before the above cited other scattering mechanism
reduce the conductivity. This behavior is in complete analogy
to the one observed for the shear viscosity $\eta(T)$ of
Helium 3 in the B phase \cite{vollh90b}.
In this system like in the system of spinless fermions
in the AB-CDW one observes an exponentially diverging
mean free path since the collision between (quasi-)particles is
suppressed by a gap. In the so-called ``Knudsen regime''
collisions of quasi-particles with the wall
of the container dominate the collisions {\em between}
the quasi-particles.
In Helium 3, one observes a sharp drop below $T_c$
and then the beginning of a plateau
before finally $\eta(T)$ vanishes rapidly.
The theoretical result for the infinite
system predicts a gentle uprise just like the one
we predict in fig.\ \ref{fi:26}. In both cases, a
factor $\exp(-\omega_\Delta/T)$ in the DOS cancels
with the same factor in the scattering rate \cite{vollh90b}.
This interesting analogy  underlines the validity of the
results of our $1/d$ approach.

\section{Discussion}
Two main questions are addressed in the present paper:
(i) How one can an infinite
dimensional result be improved by including $1/d$ corrections
in a systematic way? (ii) Which influence does spontaneous
symmetry breaking have on the conductivity?

It turned out that it is highly non trivial to
construct systematic and reasonable  approximations to arbitrary
order. This is true already on the conceptual level.
It was argued in detail
that the self-consistent calculation has certain advantages
since it yields thermodynamically consistent and conserving
approximations. The Baym/Kadanoff formalism, however, is
{\em not sufficient} to guarantee an approximation
which is free from obvious
contradictions. It was shown that an inappropriate
approximation may lead to the wrong analytic behavior
of Green functions and self-energies even though the
approximation was derived from a generating functional.

A general theorem was presented which allows to judge
whether wrong analyticity may occur. If
the conditions of the theorem are fulfilled
the appearance of the wrong analyticity is excluded.
This theorem explains a couple of observations
which were made in the last years on the application of
perturbation expansions and/or $1/d$ expansions.
It is  used to show that the self-consistent
treatment of $1/d$ corrections for spinless fermions
is a good  approximation: it possesses the necessary
analytic behavior.

For $1/d$ corrections in the Hubbard model the presented
theorem makes no statement since the self-energy has already
an imaginary part for $d=\infty$. This does not imply that the
systematic inclusion of $1/d$ corrections for the Hubbard model
is impossible, but one may expect further
difficulties. As a matter of fact, analyticity problems have been
encountered in the first calculations of $1/d$ corrections in the 
Hubbard model \cite{georg96}.

It should be stated that the self-consistent
treatment of $1/d$ corrections to any finite order in $1/d$
remains a mean-field theory. As in the $d=\infty$ treatment of
the Hubbard model \cite{janis92a} the mean field is dynamic, i.e.\
it retains a dependence on frequency.
But in the skeleton diagrams, which are considered
in any finite order in $1/d$, only lattice sites of
{\em finite} distance occur. This means that critical
fluctuations are always cut off. In $d=1$, for instance, the inclusion
of $1/d$ corrections reduces the order parameter considerably
\cite{halvo94} but does not destroy the order completely.

In the self-consistent $1/d$ treatment of spinless
fermions, two-particle properties can be reached, too.
In this work, the Bethe-Salpeter equation was set up in general and
solved in the particular case of the conductivity
$\sigma(\omega)$. This was possible for the non symmetry
broken phase as well as for the charge density wave.
The equations were evaluated in $d=3$ since the approximation
should yield the best results for this value of all
experimentally accessible dimensions \cite{halvo94}.

A number of phenomena were described in the $1/d$
expansion which can be compared with other theoretical
predictions or experiments:\\
-- the dynamic conductivity $\sigma(\omega)$ in the
homogeneous phase has a Drude peak. Its width
decreases quadratically in $T$ for small values of $T$.
The dc-conductivity is always finite \cite{uhrig95a}.\\
-- The Drude peak persists in the CDW but its weight
vanishes exponentially $\propto \exp(-\omega_\Delta/T)$, where
$\omega_\Delta$ is the gap in the one-particle spectra.
The height of the Drude peak, however, does {\em not} vanish
since the diverging quasi-particle life-time cancels the
vanishing density of states.\\
-- The real part of $\sigma(\omega)$ displays a band edge
at $\approx 2\omega_\Delta$. The singularity at the edge
is a square root. Just below the edge an excitonic
resonance is situated which is the bound state between a
particle and a hole in the empty and in the full band, respectively.
These bands are created by the spontaneous symmetry breaking.\\
-- For strong interactions the real part of $\sigma(\omega)$
shows exponentially decreasing peaks at
$\omega \approx 2m \omega_\Delta ; m\in \{1,2,3,\ldots  \}$,
which reflect the peaks in the one-particle DOS
at $\omega \approx (2m -1) \omega_\Delta$.\\
-- Strictly speaking, there is no metal-insulator transition.
But the Drude weight decays rapidly on $T\to 0$. Finally,
other scattering mechanisms will dominate over
quasi-particle--quasi-particle collisions.

In summary, we conclude that the self-consistent treatment
of $1/d$ corrections describes successfully a large
variety of phenomena since it includes the leading frequency
dependence of the self-energy. It is a generalized and improved
mean-field theory.

\section*{Acknowledgements:}
The author is grateful to D.~Vollhardt
 and E.~M\"uller-Hartmann for valuable hints and to Th.\ Pruschke
for the data shown in the inset of fig.\ \ref{fi:res1}.
The author would like to
thank H.~J.~Schulz, V.~Jani\v{s}, P.~G.~J.~van Dongen, 
and R.~Vlaming for helpful discussions and the
Laboratoire de Physique des Solides for its
hospitality. Furthermore,
the author acknowledges the financial support of the
Deutsche Forschungsgemeinschaft (SFB 341) and of the
European Community (grant ERBCHRXCT 940438).

\appendix

\section{General derivation of the theorem}

Let us consider a general correlated fermion problem without
magnetic field. A one-particle basis $\{ b^+_i\}$ ($b^+_i$ 
fermionic creation operator) can be chosen in which the 
free (one-particle) Hamiltonian can be represented as a real
matrix $\mbox{\boldmath ${h}$}^{(0)}$. The retarded self-energy and the 
retarded full (interacting) Green function are matrices as well.
They might be complex. The argument runs at
 an arbitrary but fixed value of $\omega$. According to the Dyson
equation one has 
\begin{equation}\label{aher1}
\mbox{\boldmath ${G}$}= \left( \omega+0i-\mbox{\boldmath ${h}$}^{(0)} 
-\mbox{\boldmath ${\Sigma}$}\right)^{-1} \ .
\end{equation}
The full Green function can be written as
 $\mbox{\boldmath ${G}$}
=\mathop{\rm Re}\nolimits \mbox{\boldmath ${G}$}+
i\mathop{\rm Im}\nolimits \mbox{\boldmath ${G}$}$,
where the real and the imaginary parts are real, symmetric matrices.
The same is true for the self-energy. Let us define in 
particular $\mbox{\boldmath ${B}$}:= -\mathop{\rm Im}
\nolimits \mbox{\boldmath ${\Sigma}$}$.

The aim is to show (a) that the leading order $\mbox{\boldmath ${B}$}^{(m)}$
in $\lambda$ of the matrix $\mbox{\boldmath ${B}$}$
 is positive semi-definite
and (b) that this implies that all spectral densities
are non negative. The main difference to the argument in the
main part is that the matrices do not commute
in general.

All one-particle Green functions $-<{\cal T}\{a(t)a^+(\gamma)\}>$
have positive spectral densities \cite{ricka80}. Here
 ${\cal T}$ is the time ordering operator and
$a^+$ ($a$) is an arbitrary fermionic creation (annihilation)
operator. Hence, one has for the corresponding vector ${\bf v}$
defined by $a^+ =: \sum_i v_i b^+_i$
\begin{equation}\label{aher2}
0 \leq {\bf v}^+ 
\left(\mathop{\rm Im}\nolimits \mbox{\boldmath ${G}$}\right) {\bf v} \ .
\end{equation}
Since the above equation holds for any ${\bf v}$, it implies that
the imaginary part of $\mbox{\boldmath ${G}$}$ is negative semi-definite.
Expanding (\ref{aher1}) in powers of $\mbox{\boldmath ${B}$}$
 and resumming the
imaginary part yields
\begin{equation}\label{aher3}
\mathop{\rm Im}\nolimits \mbox{\boldmath ${G}$} = -
\mbox{\boldmath ${A}$}^{-1}\mbox{\boldmath ${B}$}\,
 \mbox{\boldmath ${A}$}^{-1}
\left(1 + \mbox{\boldmath ${B}$}\,
\mbox{\boldmath ${A}$}^{-1}\mbox{\boldmath ${B}$}\,
 \mbox{\boldmath ${A}$}^{-1} \right) \ ,
\end{equation}
where $\mbox{\boldmath ${A}$}:=\omega-
\mbox{\boldmath ${h}$}^{(0)}-\mathop{\rm Re}\nolimits 
\mbox{\boldmath ${\Sigma}$}$. In leading
order in $\lambda$ this becomes
\begin{equation}\label{aher4}
\mathop{\rm Im}\nolimits \mbox{\boldmath ${G}$} = -
\lambda^m (\omega-\mbox{\boldmath ${h}$}^{(0)})\mbox{\boldmath ${B}$}^{(m)}
(\omega-\mbox{\boldmath ${h}$}^{(0)})
+{\cal O}(\lambda^{(m+1)})\ .
\end{equation}
The negative semi-definitenes of the l.\ h.\ s.\ of (\ref{aher4})
implies the positive semi-definiteness of $\mbox{\boldmath ${B}$}$ for values
of $\omega$ which are no eigenvalues of $\mbox{\boldmath ${h}$}^{(0)}$.
Assuming continuity \cite{note2}
 for $\mbox{\boldmath ${B}$}^{(m)}(\omega)$ 
the positive semi-definiteness extends to all frequencies.

Addressing the sign of the imaginary part of the full Green 
function we state that $\mbox{\boldmath ${B}$}\geq0$ implies that there
is a matrix $\sqrt{\mbox{\boldmath ${B}$}}$ which is real and symmetric
as well. Defining 
$\mbox{\boldmath ${D}$}:= \sqrt{\mbox{\boldmath ${B}$}}\, 
\mbox{\boldmath ${A}$}^{-1}\sqrt{\mbox{\boldmath ${B}$}}$ 
allows (\ref{aher3}) to be written as
\begin{equation}\label{aher5}
\mathop{\rm Im}\nolimits \mbox{\boldmath ${G}$} = -
\mbox{\boldmath ${A}$}^{-1}\sqrt{\mbox{\boldmath ${B}$}} 
\left(1+\mbox{\boldmath ${D}$}^2\right)^{-1}  
\sqrt{\mbox{\boldmath ${B}$}}\, \mbox{\boldmath ${A}$}^{-1}\ .
\end{equation}
The expression in parentheses is manifestly positive semi-definite,
thus its inverse as well. 
Since $\mbox{\boldmath ${A}$}^{-1}\sqrt{\mbox{\boldmath ${B}$}}$ 
and $\sqrt{\mbox{\boldmath ${B}$}}\, \mbox{\boldmath ${A}$}^{-1}$
are transposed to each other, the r.\ h.\ s.\ in (\ref{aher5})
is negative semi-definite as a whole. This concludes the argument.

\section{Conductivity in the AB-CDW}
In this appendix the geometric series of fig.\ \ref{fi:21}
is derived in the case  of finite order parameter. Furthermore 
the derivation of (\ref{wfsum0}) is given.

\begin{figure}[hbt]
\setlength{\unitlength}{1cm}
\begin{picture}(8.2,3.3)(0,0)
\put(12.5,0.6){\psfig{file=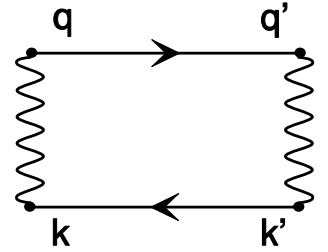,height=3.3cm,width=4cm}}
\end{picture}
\caption{Diagrammatic basis element of the terms in the
	sum in fig.\ \protect\ref{fi:21}}
	\label{fi:c1}
\end{figure}

In the AB-CDW the propagators are not diagonal in ${\bf k}$-space.
There is the possibility of a transition ${\bf k} \to {\bf k} + {\bf Q}$.
The matrix element for this process is the off-diagonal element
in (\ref{block2}). If the wave vector remains unchanged
the diagonal matrix elements have to be used. Let us 
classify the segments between two adjacent wavy lines in a
diagram in fig.\ \ref{fi:21}. A generic segment is shown in
 fig.\ \ref{fi:c1}.
 Note that the momentum of the upper propagator cancels the
momentum in the lower propagator up to multiples of 
${\bf Q}=(\pi,\pi,\pi)^\dagger$, since we are interested in the
average conductivity. At the end-vertices no momentum is
added or subtracted. The case ${\bf k} = {\bf k}' = {\bf q} = {\bf q}'$ 
can be combined with the case
${\bf k} = {\bf q} = {\bf k}'+{\bf Q} = {\bf q}'+{\bf Q}$.
In both cases the momentum through the whole segment is zero.
Using the elements from (\ref{block2}) one obtains
\begin{equation}
\label{fall1}
A_1 = -\frac{2UT}{Z}
\sum\limits_{\omega_\nu-\omega_\lambda=\omega_m}
\; \int\limits_{-\infty}^\infty \frac{w_\nu w_\lambda+(\gamma\varepsilon)^2
-\Delta(i\omega_\nu)\Delta(i\omega_\lambda)}
{\left(w_\nu^2- (\gamma\varepsilon)^2-\Delta^2(i\omega_\nu)\right)
\left(w_\lambda^2- (\gamma\varepsilon)^2-\Delta^2(i\omega_\lambda)\right)}
N_{c,0}(\varepsilon)d\varepsilon
\ ,\qquad\quad\end{equation}
where the short-hand $w_{\nu/\lambda}
:=i\omega_{\nu/\lambda}-\Sigma(i\omega_{\nu/\lambda})$ is employed again.
A similar expression is obtained in the combined cases
${\bf k} = {\bf k}' = {\bf q}+{\bf Q} = {\bf q}'+{\bf Q}$ and
 ${\bf k} = {\bf q}+{\bf Q} = {\bf k}'+{\bf Q} = {\bf q}'$ for which
the momentum through the segment is ${\bf Q}$
\begin{equation}
\label{fall2}
A_2 =-\frac{2UT}{Z}
\sum\limits_{\omega_\nu-\omega_\lambda=\omega_m}
\; \int\limits_{-\infty}^\infty \frac{w_\nu w_\lambda-(\gamma \varepsilon)^2
-\Delta(i\omega_\nu)\Delta(i\omega_\lambda)}
{\left(w_\nu^2- (\gamma\varepsilon)^2-\Delta^2(i\omega_\nu)\right)
\left(w_\lambda^2- (\gamma\varepsilon)^2-\Delta^2(i\omega_\lambda)\right)}
N_{c,0}(\varepsilon)d\varepsilon
\ .\qquad\quad\end{equation}
The segments which {\em change} the momentum are very important.
The cases 
${\bf k} = {\bf k}' = {\bf q}+{\bf Q} = {\bf q}'$ and
 ${\bf k} = {\bf q}+{\bf Q} = {\bf k}'+{\bf Q} = {\bf q}'+{\bf Q}$
yield together
\begin{equation}
\label{fall3}
A_3 =
-\frac{2UT}{Z}\sum\limits_{\omega_\nu-\omega_\lambda=\omega_m}
\; \int\limits_{-\infty}^\infty \frac{w_\nu \Delta(i\omega_\lambda)
-w_\lambda\Delta(i\omega_\nu)}
{\left(w_\nu^2- (\gamma\varepsilon)^2-\Delta^2(i\omega_\nu)\right)
\left(w_\lambda^2- (\gamma\varepsilon)^2-\Delta^2(i\omega_\lambda)\right)}
N_{c,0}(\varepsilon)d\varepsilon
\ .\qquad\quad\end{equation}

The possible combinations of these cases are naturally generated
by powers of the matrix
\begin{equation}\label{fall4}
\mbox{\boldmath ${A}$} := \left( 
\begin{array}{ccc}
A_1 & A_3 \\
A_3 & A_2
\end{array} \right)\ .
\end{equation}
The sum of all these powers is a geometric series yielding
finally
\begin{equation}\label{fall5}
\chi^{\scriptstyle \rm JJ} = 
-\frac{2}{U} \left(\left(1-\mbox{\boldmath ${A}$}\right)^{-1}
\right)_{(1,1)}
\ .\end{equation}
Taking the (1,1) element ensures that the average current-current
correlation function is calculated. The prefactor compensates the fact that
the end vertices do not have the factor $-U/2$ which is incorporated
in $\mbox{\boldmath ${A}$}$. Eq.\ (\ref{vollchi2}) is a direct consequence of 
(\ref{fall5}). The representation (\ref{basis0}) follow from 
(\ref{fall1}), (\ref{fall2}), and (\ref{fall3}) by partial fraction
decomposition and integration over the conductivity spectral density
using the conductivity Green functions.

Now we turn to (\ref{wfsum0}).  For $\chi^{\scriptstyle \rm JJ}(0)$ we need
only $A_1$ at $i\omega_m=0$ since $A_3$ vanishes at $i\omega_m=0$
and the factor $1-A_2$ then drops out. For $A_1$ one obtains
\begin{mathletters}\label{wfsum1}
\begin{eqnarray}
A_1 &=& -\frac{2UT}{Z}\sum_\nu
\int\limits_{-\infty}^\infty\frac{(w_\nu+\gamma\varepsilon)^2-
\Delta^2(i\omega_\nu)}
{\left(w_\nu^2-(\gamma\varepsilon)^2-\Delta^2(i\omega_\nu)\right)^2}
N_{c,0}(\varepsilon)d\varepsilon
\\ \label{wfsum1b}
&=&  -\frac{2UT}{Z}\sum_\nu
\int\limits_{-\infty}^\infty\frac{w_\nu^2-
(\gamma\varepsilon)^2-\Delta^2(i\omega_\nu)
+2(\gamma\varepsilon)^2}
{\left(w_\nu^2-(\gamma\varepsilon)^2-\Delta^2(i\omega_\nu)\right)^2}
N_{c,0}(\varepsilon)d\varepsilon
\\ \label{wfsum1c}
&=& -\frac{2UT}{Z}\sum_\nu
\int\limits_{-\infty}^\infty\frac{N_{c,0}(\varepsilon) -
\left.\partial \left(\varepsilon N_{c,0}(\varepsilon) \right)\right/
\partial \varepsilon}
{w_\nu^2-(\gamma\varepsilon)^2-\Delta^2(i\omega_\nu)}d\varepsilon
\\ \label{wfsum1d}
&=& -\frac{UT}{Z}\sum_\nu
\int\limits_{-\infty}^\infty\frac{\varepsilon^2}
{w_\nu^2-(\gamma\varepsilon)^2-\Delta^2(i\omega_\nu)} N_0(\varepsilon)
d\varepsilon
\\ 
&=& \frac{UT}{\pi Z} 
\int\limits_{-\infty}^\infty f_{\scriptstyle \rm F}(\omega)
 \mathop{\rm Im}\nolimits
\int\limits_{-\infty}^\infty\frac{\varepsilon^2 
N_0(\varepsilon)d\varepsilon d\omega}
{(\omega-\Sigma(\omega))^2-(\gamma\varepsilon)^2-\Delta^2(\omega)}
\ .\qquad\quad
\end{eqnarray}\end{mathletters}
The step to (\ref{wfsum1b}) follows from the fact that 
$N_{c,0}(\varepsilon)$ is even.
Partial integration of the last term of the numerator 
leads to  (\ref{wfsum1c}). 
Eq.\ (\ref{wfsum1d}) is obtained using (\ref{zufo3}) which emphasizes
the importance of this relation between DOS and conductivity
spectral density. In the end one obtains
\begin{mathletters}\label{wfsum2}
\begin{eqnarray}
A_1 &=& -\frac{\sqrt{Z}\Sigma^{\scriptstyle \rm F}}{\gamma}
\\ \label{wfsum2b}
&=& -\frac{U\langle \hat{T}\rangle}{ \gamma Z}
\\ \label{wfsum2c}
&=& -(1/\gamma-1)
\ .\end{eqnarray}\end{mathletters}
One obtains (\ref{wfsum0}) by substituting (\ref{wfsum2c})
in the denominator of (\ref{vollchi2}) and comparing the
resulting expression with (\ref{wfsum2b}). This completes
the analytic derivation of the f-sum rule in the self-consistent
$1/d$ approximation which is considered here.

\newpage

\begin{thebibliography}{10}

\bibitem{metzn89a}
W.~Metzner and D.~Vollhardt, Phys. Rev. Lett. {\bf 62}, 324 (1989)

\bibitem{vollh93}
D.~Vollhardt, in {\em Correlated Electron Systems}, by V.~J. Emery, p.~57
  (World Scientific, Singapore, 1993)

\bibitem{georg96}
A.~Georges, G.~Kotliar, W.~Krauth and M.~J. Rozenberg, 
Rev. Mod. Phys. {\bf 68} 13 (1996)

\bibitem{prusc96}
Th. Pruschke, M.~Jarrell and J.~K. Freericks, 
Adv. Phys. {\bf 44}, 187 (1995)

\bibitem{mulle89a}
E.~M\"uller-Hartmann, Z. Phys. B {\bf 74}, 507 (1989)

\bibitem{halvo94}
E.~Halvorsen, G.~S. Uhrig and G.~Czycholl, Z. Phys. B {\bf 94}, 291 (1994)

\bibitem{strac91}
R.~Strack and D.~Vollhardt, J. Low Temp. Phys. {\bf 84}, 357 (1991)

\bibitem{schwe91a}
H.~Schweitzer and G.~Czycholl, Z. Phys. B {\bf 83}, 93 (1991)

\bibitem{janis92a}
V.~Jani\v{s} and D.~Vollhardt, Int. J. Mod. Phys. {\bf 6}, 731 (1992)

\bibitem{ricka80}
G.~Rickayzen, {\em Green's Functions and Condensed Matter} (Academic Press,
  London, 1980)

\bibitem{schwe90b}
H.~Schweitzer and G.~Czycholl, Z. Phys. B {\bf 79}, 377 (1990)

\bibitem{baym61}
G.~Baym and L.~P. Kadanoff, Phys. Rev. {\bf 124}, 287 (1961)

\bibitem{baym62}
G.~Baym, Phys. Rev. {\bf 127}, 1391 (1962)

\bibitem{schwe91b}
H.~Schweitzer and G.~Czycholl, Phys. Rev. Lett. {\bf 67}, 3724 (1991)

\bibitem{uhrig95d}
G.~S. Uhrig and R.~Vlaming, Ann. Physik {\bf 4}, 778 (1995)

\bibitem{note1}
The imaginary part is in fact antisymmetric in $\omega$ so that
one obtains a contradiction by choosing an appropriate sign
of $\omega$, too.

\bibitem{schwe90a}
H.~Schweitzer and G.~Czycholl, Solid State Commun. {\bf 74}, 735 (1990)

\bibitem{schil95}
A.~Schiller and K.~Ingersent, Phys. Rev. Lett. {\bf 75}, 113 (1995)

\bibitem{mulle89b}
E.~M\"uller-Hartmann, Z. Phys. B {\bf 76}, 211 (1989)

\bibitem{menge91}
B.~Menge and E.~M\"uller-Hartmann, Z. Phys. B {\bf 82}, 237 (1991)

\bibitem{lutti61}
J.~M. Luttinger, Phys. Rev. {\bf 121}, 942 (1961)

\bibitem{donge91}
P.~G.~J. van Dongen, Phys. Rev. Lett. {\bf 67}, 757 (1991)

\bibitem{donge94b}
P.~G.~J. van Dongen, Phys. Rev. B {\bf 50}, 14016 (1994)

\bibitem{mahan90}
G.~D. Mahan, {\em Many-Particle Physics -2nd ed.} (Plenum Press, New York,
  1990)

\bibitem{khura90}
A.~Khurana, Phys. Rev. Lett. {\bf 64}, 1990 (1990)

\bibitem{uhrig95a}
G.~S. Uhrig and D.~Vollhardt, Phys. Rev. B {\bf 52}, 5617 (1995)

\bibitem{uhrig95c}
G.~S. Uhrig, Physica B {\bf 206/207}, 698 (1995)

\bibitem{prusc93a}
Th. Pruschke, D.~L. Cox and M.~Jarrell, Europhys. Lett. {\bf 21}, 593 (1993)

\bibitem{prusc93b}
Th. Pruschke, D.~L. Cox and M.~Jarrell, Phys. Rev. B {\bf 47}, 3553 (1993)

\bibitem{vollh90b}
D.~Vollhardt and P.~W\"olfle, {\em The Superfluid Phases of Helium 3} (Taylor
  and Francis, London, 1990)

\bibitem{note2}
The argument is applied first to finite systems for which the
Hilbert space is finite dimensional and the spectra discrete.
The theorem's statement extends to the thermodynamic limit if
this limit exists.
\end{thebibliography}

\end{document}